\documentclass[10pt]{article}

\usepackage[english]{babel}
\usepackage[letterpaper,top=2.5cm,bottom=2.5cm,left=2.5cm,right=2.5cm]{geometry}
\usepackage{amsmath}
\usepackage{graphicx}
\usepackage{subcaption}
\usepackage{booktabs}
\usepackage{xcolor}
\usepackage{textcomp}
\usepackage{multirow}
\usepackage{amssymb}
\usepackage[colorlinks=true, allcolors=blue]{hyperref}
\usepackage{lineno}
\usepackage{tikz}

\newcommand{\cmark}{\checkmark}
\newcommand{\xmark}{\texttimes}

\title{Fine-tuning MLIP foundation models: strategies for accuracy and transferability}

\author{
Tam\'as Lajos Tompa$^{\dagger}$,
Eszter Varga-Umbrich$^{1,\dagger}$,
Ilyes Batatia$^{1}$ \\
Alin M. Elena$^{2}$,
Noam Bernstein$^{3}$,
G\'abor Cs\'anyi$^{1}$ \\[0.5em]
\small $^{1}$Department of Engineering, University of Cambridge, Cambridge, UK \\
\small $^{2}$Scientific Computing Department, Science and Technology Facilities Council--Daresbury Laboratory, Warrington, UK \\
\small $^{3}$Center for Materials Physics and Technology, U.S. Naval Research Laboratory, Washington, DC, USA \\
\small $^{\dagger}$These authors contributed equally. \\
}

\date{}

\begin{document}
\maketitle
\begin{abstract}

Adapting machine-learned interatomic potential (MLIP) foundation models to specialised tasks through fine-tuning is an increasingly important practice, yet systematic guidance on when and how to fine-tune is currently limited. We evaluate seven fine-tuning strategies---naive full-parameter updates, two layer-freezing variants, Low-Rank Adaptation (LoRA), multihead replay, pseudolabelled replay, and replay combined with LoRA---across five chemically diverse benchmarks (aqueous NaCl, ice polymorphs, S$_\mathrm{N}$2 reactions, SPICE biomolecules, and lithium electrolytes), three generations of foundation models, and training sets spanning five orders of magnitude. To support this evaluation we implement three capabilities in the MACE codebase: LoRA adapted for equivariant message-passing architectures, including both scalar and equivariant linear layers; pseudolabelled replay, which decouples the replay data source from the original pretraining corpus; and model-aware atomic reference energy (E0) reestimation for fine-tuning workflows. We find that foundation model quality, correct E0 initialisation, and well-chosen hyperparameters are prerequisites whose impact routinely exceeds that of the fine-tuning strategy itself. Once these prerequisites are met, most strategies achieve strong target-task accuracy, consistently surpassing models trained from scratch. The practical distinction depends on deployment scope: naive fine-tuning offers the best convergence for single-system applications, while multihead replay---with either original or pseudolabelled data---is the only approach tested that consistently preserves out-of-distribution robustness, maintaining both pretraining-distribution accuracy for broader deployment and many-body short-range repulsion.
\end{abstract}

\section{Introduction}

Foundation models for machine-learned interatomic potentials (MLIPs)~\cite{mace_mp0,chgnet, uma, gnome, orb, sevennet, mh-1} have demonstrated transferability across various chemical systems. This generalisation capability enables a fundamentally new workflow for modelling materials with MLIPs: rather than training task-specific potentials from scratch---a process requiring extensive human and computational resources for data generation~\cite{unke_review_2021, behler_review_2021, smith_anial_2018, schran_committee_al}---practitioners can adapt pretrained foundation models to specialised applications through fine-tuning~\cite{mace_mp0, kaur2024dataefficientfinetuningfoundationalmodels, Della_Pia_2025, PETMADLora, macefreeze, Wang_2025, noam_phonon}. Reported successes span data-efficient specialisation to narrowly defined chemical systems with a small fraction of the configurations required by training from scratch~\cite{kaur2024dataefficientfinetuningfoundationalmodels}, parameter-efficient adaptation that keeps the additional memory and compute cost modest~\cite{PETMADLora}, and targeted updates that retain most of the pretrained behaviour while improving accuracy on the system of interest~\cite{macefreeze}. Across these settings fine-tuned models routinely match or exceed bespoke from-scratch potentials at the target task while requiring orders of magnitude less reference data, making fine-tuning an increasingly default starting point for system-specific MLIP development rather than a niche option~\cite{mace_mp0}.

However, early reports of fine-tuning raised concerns about catastrophic forgetting~\cite{catastrophic_forgetting, kaur2024dataefficientfinetuningfoundationalmodels, Della_Pia_2025, mace_mp0}, prompting the adoption of constrained fine-tuning techniques originally developed for deep, unstructured networks---including Low-Rank Adaptation (LoRA)~\cite{lora} and its adaptations to atomistic neural networks, both equivariant~\cite{eloraGNN} and non-equivariant~\cite{PETMADLora, eloraGNN}, experience replay implemented via multihead architectures~\cite{mace_mp0}, and weight freezing to preserve learned representations~\cite{macefreeze}---implicitly treating MLIP fine-tuning as prone to the same fragility as convolutional neural networks (CNNs) or large language models (LLMs).

\begin{figure}[h]
    \centering
    \includegraphics[width=1\linewidth]{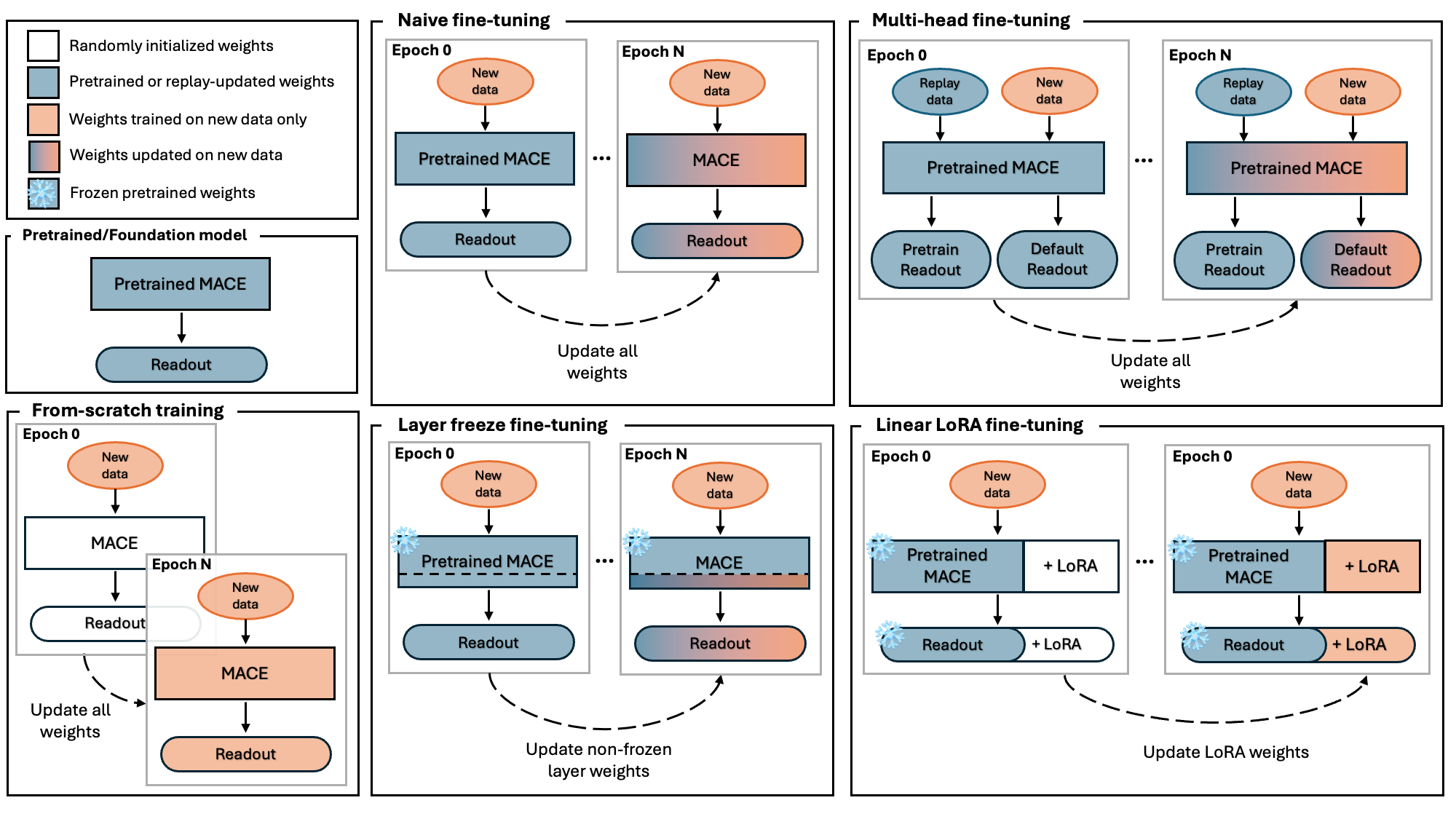}
    \caption{\textbf{Overview of fine-tuning strategies evaluated.} Schematic representation of from-scratch model training and the fine-tuning methods benchmarked in this work: naive fine-tuning (full parameter updates), layer freezing, low-rank adaptation (LoRA), multihead replay, pseudolabelled replay, and replay combined with LoRA.
    }
    \label{fig:fine-tuning}
\end{figure}
Here we hypothesise that early fine-tuning failures were primarily the result of weaker foundation models, improper atomic reference energy initialisation, and destabilising training procedures. While forgetting remains a concern in some settings, as demonstrated later, these factors have likely overstated the intrinsic limitations of naive fine-tuning. We therefore set out to characterise the major factors shaping fine-tuning outcomes and how each affects target-task accuracy and out-of-distribution behaviour. To do so we evaluate seven fine-tuning strategies: naive fine-tuning (continued gradient descent on all model parameters from the foundation checkpoint), two layer-freezing variants, LoRA at multiple ranks, multihead replay with original labels, multihead replay with pseudolabels, and replay combined with LoRA (Fig.~\ref{fig:fine-tuning}). We apply them across a suite of benchmark systems chosen to span increasing departure from OMat24's core domain of periodic inorganic bulk materials: lithium argyrodite electrolytes, ice polymorphs, aqueous NaCl~\cite{oneill2024pairpairmachinelearnedexplicitlycorrelated}, S$_\mathrm{N}$2 reactions~\cite{kuryla2025efficient}, and biomolecules from the SPICE dataset~\cite{spice} (Fig.~\ref{fig:benchmark_systems}); we additionally test on metal--organic framework configurations~\cite{Elena2025} in the context of universal-model isolated-atom behaviour. Together these cover inorganic and hybrid metal--organic crystalline solids, molecular solids, ionic solutions, gas-phase reactions, and organic and biomolecular conformers. Our evaluation spans five orders of magnitude in training set size (5 to 950k configurations), several MACE foundation checkpoints and architectures (including MACE-MP0a-medium, MACE-OMat-0-small, MACE-OMat-0-medium and MACE-MH1), and systematic comparisons between recent OMat24-based~\cite{omat24} and earlier MPTraj-based~\cite{chgnet, mace_mp0} foundation models.

Crucially, we go beyond static, pointwise energy and force errors and probe dynamic and extrapolative behaviour: radial distribution functions from MD trajectories, NEB reaction profiles~\cite{neb}, MD stability tests, and random structure search (RSS)~\cite{airss_pickard} to detect failures of the short-range repulsion. These protocols probe different aspects of the learned potential energy surface (PES) and reveal failure modes that pointwise validation alone would miss. Detailed descriptions of each protocol are given in Methods.

\section{Methods}

Throughout this work we fine-tune MACE foundation models pretrained on inorganic datasets (OMat24 and MPTraj). 
We choose these models because they are applicable to the broadest range of chemical systems, enabling a more complete picture of fine-tuning across diverse chemistry. Importantly, MACE models trained on inorganic crystals have been shown to generalise broadly across chemistry~\cite{mace_mp0}, and we deliberately include benchmark tasks not just in periodic structures, but in organic, molecular, and reactive regimes where one might not a priori expect an inorganic-pretrained model to excel, probing whether learned representations are transferable, and how different fine-tuning strategies affect it.

\subsection{Benchmark systems}
\label{sec:benchmarks}
\begin{figure}[h]
    \centering
    \begin{tikzpicture}
        \def\imgwidth{\linewidth}   
        \def\cropheight{7.5cm}       
        \begin{scope}
            \clip (0,0) rectangle (\imgwidth,\cropheight);
            \node[inner sep=0pt, anchor=north west] (bench) at (0,\cropheight)
                {\includegraphics[width=\imgwidth]{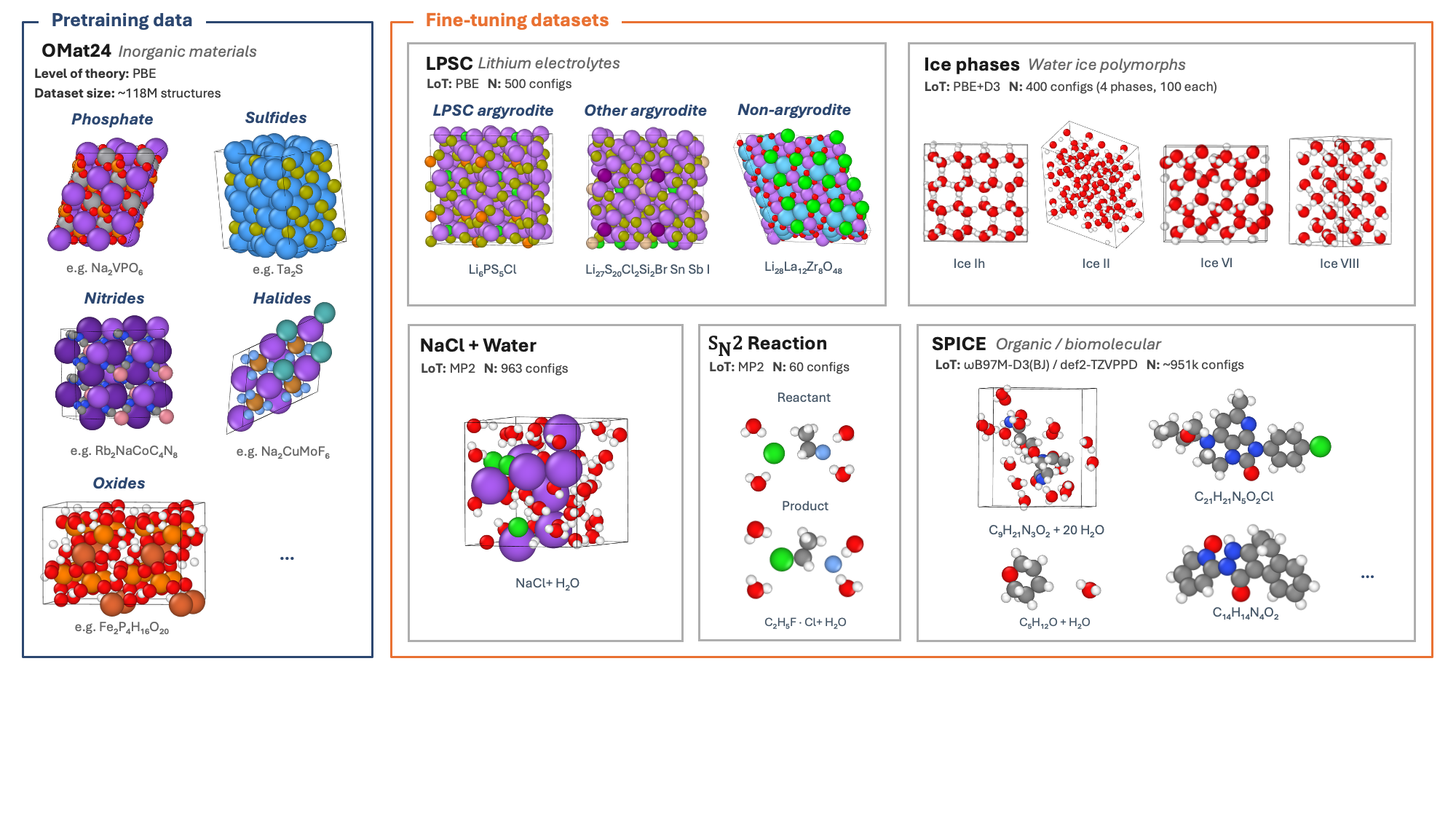}};
        \end{scope}
    \end{tikzpicture}
    \caption{\textbf{Overview of the five main benchmark systems used in this work.} Representative structures and key dataset attributes for each benchmark, ordered by increasing departure from OMat24's core domain of periodic inorganic bulk materials: the argyrodite Li$_6$PS$_5$Cl is an inorganic periodic solid most aligned with the pretraining data, while aqueous NaCl, ice polymorphs, S$_\mathrm{N}$2 reactions, and SPICE biomolecules progressively probe further from this domain. Detailed reference levels, dataset sizes, evaluation sets, and replay-set composition are given in Appendix Table~\ref{tab:benchmark_systems}.}
    \label{fig:benchmark_systems}
\end{figure}

We use five benchmark systems chosen to span increasing departure from OMat24's core domain of periodic inorganic bulk materials (Fig.~\ref{fig:benchmark_systems}). This spread is what allows us to separate fine-tuning methods: on systems already close to the pretraining manifold, the foundation model encodes most of the relevant chemistry and all methods tend to perform similarly, whereas systems further from the pretraining domain expose method-level differences in target-task accuracy, transfer, and forgetting. At one end of this spread, the lithium electrolyte Li$_6$PS$_5$Cl (LPSC) is an inorganic periodic solid of exactly the kind that is common in OMat24, so we expect methods to be hard to distinguish when fine-tuning only on this system. At the other end, SPICE biomolecules and S$_\mathrm{N}$2 reactions probe organic, molecular, and reactive chemistry entirely absent from the pretraining corpus, which contains no gas-phase molecules. Aqueous NaCl and ice polymorphs sit in between, adding solvation structure and molecular-solid phase behaviour. Reference levels of theory, dataset sizes, evaluation sets, and replay-set composition are summarised in Appendix Table~\ref{tab:benchmark_systems}.

\textbf{Lithium electrolytes.} Fine-tuning uses 500 PBE-reference configurations of a single argyrodite composition, Li$_6$PS$_5$Cl. Evaluation is performed on a broad set of \emph{other} lithium electrolytes never seen during fine-tuning: over 130 argyrodite-structured compositions other than LPSC, plus non-argyrodite lithium-electrolyte structures whose elemental compositions can differ from the LPSC training composition apart from the shared presence of Li. This probes out-of-distribution compositional and structural transfer within a chemical family the foundation model is already well suited for.

\textbf{Aqueous NaCl.} A 963-configuration MP2-reference dataset~\cite{oneill2024pairpairmachinelearnedexplicitlycorrelated}, with subsampling used for data-efficiency experiments. Evaluation uses radial distribution functions from MD trajectories, compared against the Behler--Parrinello neural network (BPNN) reference from the same work, providing a structural metric beyond pointwise energy and force errors. While the BPNN reference is itself a fitted potential rather than a direct experimental measurement, it was validated against MP2 energies and forces and reproduces known solvation structure, making it a suitable structural benchmark for this comparison.

\textbf{Ice polymorphs.} 400 PBE+D3 configurations (100 per phase) across ice Ih, II, VI, and VIII~\cite{kaur2024dataefficientfinetuningfoundationalmodels}, with separate held-out test configurations per phase. Cross-phase force errors and MD stability tests probe transfer within a related molecular-solid family.

\textbf{S$_\mathrm{N}$2 reactions.} 60 MP2-reference configurations covering reactant and product states only, excluding the transition state~\cite{kuryla2025efficient}. Evaluation requires the model to relax NEB paths along a physically meaningful minimum-energy path rather than evaluating a fixed reference trajectory, with barrier heights compared against the MP2 reference. This couples narrow training data to a demanding extrapolation task.

\textbf{SPICE biomolecules.} The cleaned and filtered subset used to fit MACE-OFF~\cite{spice,MACEOFF}, comprising up to 855\,905 training / 95\,100 validation configurations at $\omega$B97M-D3(BJ)/def2-TZVPPD and a separate 50\,000-configuration test set. Although SPICE introduces no new atomic species beyond those in OMat24, its covalent, conformational, and non-covalent interactions differ substantially from the foundation model's inorganic-solid pretraining domain, making it the most demanding test of both target accuracy and forgetting in our benchmark suite.

\subsection{Naive fine-tuning}

Naive fine-tuning initialises all model parameters from the pretrained checkpoint and then continues training with stochastic gradient descent on the target dataset, without architectural constraints or additional regularisation. The optimiser hyperparameters are, however, adjusted relative to from-scratch training: we use a reduced learning rate and an increased exponential moving average (EMA) decay to limit how far the model can drift from its pretrained solution within a single step. These adjustments are not regularisers in the explicit sense, but in our experiments they are essential for stable, well-converged naive fine-tuning (see Section~\ref{sec:hypers}). The method is computationally efficient, conceptually simple, and serves as our primary baseline.

\subsection{Layer freezing}
\label{sec:layer_freezing}

Layer freezing selectively updates a subset of model parameters while holding the rest fixed at their pretrained values~\cite{macefreeze}. We explore two configurations motivated by the hierarchical structure of MACE:

\textbf{Freeze embedding and message-passing layers, train readout.} All message-passing layers, tensor product operations, and embedding networks remain frozen at pretrained weights; only the final readout layers that map learned representations to energies are trained. This configuration assumes that pretrained geometric and chemical representations are already adequate and only the mapping to the target task's energy scale needs adaptation.

\textbf{Freeze embedding and first message-passing layer.} The embedding weights and first interaction block are frozen, while later interaction blocks and readouts are trainable. This assumes that low-level atomic feature extraction is universal but higher-level interaction patterns may need task-specific adaptation.

\subsection{Low-Rank Adaptation (LoRA)}

Low-Rank Adaptation~\cite{lora} was originally introduced as a parameter-efficient fine-tuning method for LLMs with several billion parameters, where only a small fraction of the weights are trainable to keep optimiser-state and gradient memory manageable. For MLIPs the parameter-efficiency motivation is much weaker---MACE foundation models are several orders of magnitude smaller than typical LLMs and fit comfortably in GPU memory, and the dominant cost is storage of atomic graphs rather than model weights. LoRA has nonetheless been applied to atomistic neural networks in this regime: to the non-equivariant PET architecture in PET-MAD~\cite{PETMADLora}, and to the equivariant layers of MACE in ELoRA~\cite{eloraGNN}. Our implementation differs from ELoRA in two respects: all layers---including embedding and readouts---of the model can be adapted during fine-tuning because we apply LoRA to both equivariant linear and regular scalar linear layers (ELoRA targets only the equivariant ones, and it also includes the tensor products), and within the equivariant linear layers we use an irrep-preserving activation-space bottleneck, rather than using a path-dependent adapter parametrisation. We adopt LoRA primarily as a way to match the effective capacity of the trainable subspace to the fine-tuning task, while keeping the pretrained representations as a fixed scaffold; we discuss the empirical evidence for this capacity-matching effect below.

We freeze all pretrained weights $\mathbf{W}_{\text{pretrain}}$ and inject trainable low-rank decompositions into the linear layers of the MACE architecture: both the regular (scalar) linear layers used in the model, and the equivariant linear layers that act on irreducible-representation (irrep) features between interaction blocks. The equivariant tensor product operations are left unchanged.

\textbf{Standard (scalar) linear layers.} For each targeted weight matrix $\mathbf{W} \in \mathbb{R}^{d_{\text{out}} \times d_{\text{in}}}$ in the pretrained model, the forward pass computes
\begin{equation}
\mathbf{h}_{\text{out}} = \left( \mathbf{W}_{\text{pretrain}} + \mathbf{B}\mathbf{A} \right) \mathbf{h}_{\text{in}},
\end{equation}
where $\mathbf{B} \in \mathbb{R}^{d_{\text{out}} \times r}$ and $\mathbf{A} \in \mathbb{R}^{r \times d_{\text{in}}}$ are trainable low-rank matrices with rank $r \ll \min(d_{\text{in}}, d_{\text{out}})$. The matrices are initialised as $\mathbf{A} \sim \mathcal{N}(0, \sigma^2)$ and $\mathbf{B} = \mathbf{0}$, ensuring that at initialisation $\mathbf{W}_{\text{pretrain}} + \mathbf{B}\mathbf{A} = \mathbf{W}_{\text{pretrain}}$.

\textbf{Equivariant linear layers.} An equivariant linear layer in MACE acts on a direct sum of irrep features $\bigoplus_{l} \mathbf{h}_l$, mixing channels within each irrep type but never across types. For each irrep block, a scalar weight matrix $\mathbf{W}_l \in \mathbb{R}^{C_l^{\text{out}} \times C_l^{\text{in}}}$ is applied independently, and we decompose the update on the same block structure:
\begin{equation}
\mathbf{W}_l \;\longrightarrow\; \mathbf{W}_{l,\,\text{pretrain}} + \mathbf{B}_l \mathbf{A}_l,
\qquad \mathbf{A}_l \in \mathbb{R}^{r \times C_l^{\text{in}}},\;\; \mathbf{B}_l \in \mathbb{R}^{C_l^{\text{out}} \times r},
\end{equation}
with the same rank $r$ used for every irrep. Equivalently, $\mathbf{A}$ projects each input irrep block to a low-rank irrep bottleneck and $\mathbf{B}$ projects it back to the full output multiplicity; irreps are never mixed. Because each $(\mathbf{A}_l, \mathbf{B}_l)$ pair is a scalar weight on the $l$-th irrep block, the LoRA update is itself an equivariant linear layer, so equivariance is preserved.

\textbf{Effective capacity and rank.} The rank $r$ controls the dimensionality of the trainable subspace. As a concrete reference point, rank-$4$ LoRA applied to all targeted (scalar and equivariant) linear layers of the medium-sized MACE-OMat foundation amounts to roughly $3\%$ of the model's total trainable parameter count; the per-layer overhead is harder to describe using a single number, because MACE contains layers of differing widths and irrep multiplicities, but in all cases $r$ is much smaller than the input and output multiplicities of the layer. The choice of $r$, its effect on target-task accuracy and forgetting, and the empirical evidence for the capacity-matching effect are discussed in Section~\ref{sec:method_comparison} and the Discussion.

\subsection{Multihead replay fine-tuning}

Multihead replay training~\cite{mace_mp0} enables simultaneous optimisation on the target dataset and a replay dataset, the latter chosen to represent broad chemical and structural configurations, using either a subset of the foundation model's pretraining data or another structurally diverse source. This is similar to experience-replay ideas from continual learning, where examples from previous tasks are interleaved with new-task updates to reduce forgetting~\cite{rolnick2019experience,lesort2019generative}. This provides a mechanism to maintain information about interaction trends across a wide range of chemical compositions and structures when the fine-tuning data are narrow but the intended application is broad. The architecture consists of a shared representation network with separate readout heads for the target and replay tasks. In our default workflow, replay structures are selected from the pretraining data by element matching: we retain only configurations whose constituent elements are a subset of those in the fine-tuning dataset, excluding structures that contain elements absent from the target data while allowing structures that contain only some of the target elements. Replay-set sources and sizes are summarised with the benchmark datasets in Appendix~\ref{app:benchmarks}.

The combined loss function is:
\begin{equation}
\mathcal{L}_{\text{total}} = \mathcal{L}_{\text{target}}(\boldsymbol{\theta}_{\text{shared}}, \boldsymbol{\theta}_{\text{head}}^{\text{target}};\, \mathcal{D}_{\text{target}}) + \mathcal{L}_{\text{replay}}(\boldsymbol{\theta}_{\text{shared}}, \boldsymbol{\theta}_{\text{head}}^{\text{replay}};\, \mathcal{D}_{\text{replay}})
\end{equation}
where $\boldsymbol{\theta}_{\text{shared}}$ denotes the shared embedding and message passing parameters, $\boldsymbol{\theta}_{\text{head}}^{\text{target}}$ and $\boldsymbol{\theta}_{\text{head}}^{\text{replay}}$ are the task-specific readout parameters, $\mathcal{D}_{\text{target}}$ is a batch from the fine-tuning dataset, and $\mathcal{D}_{\text{replay}}$ is a batch from the replay dataset. Each head maintains its own set of atomic reference energies (E0s), enabling different levels of theory or reference states for the target and replay data.

We apply distinct loss weights across the target and replay heads:
\begin{equation}
\mathcal{L}_{h} = \lambda_E^{h} \mathcal{L}_E^{h} + \lambda_F^{h} \mathcal{L}_F^{h} + \lambda_S^{h} \mathcal{L}_S^{h},
\qquad h \in \{\text{target}, \text{replay}\},
\end{equation}
where the stress term is included only for datasets with stress labels.

For replay data we use $\lambda_E^{\text{replay}} = 1$ and $\lambda_F^{\text{replay}} = 10$, matching the pretraining objective, while maintaining $\lambda_E^{\text{target}} = 10$ and $\lambda_F^{\text{target}} = 10$ for the target data. For datasets with stress labels, we use $\lambda_S = 1.0$ on the corresponding head. The higher energy weight on the target task prioritises energy accuracy for applications such as barrier heights and relative phase stabilities.

\textbf{Pseudolabel replay fine-tuning.} Within this multihead framework, the choice of replay-label source defines two variants, which we describe below. Both variants share the same architecture, loss function, and hyperparameters.
In original-label replay, the selected replay configurations are sampled from the foundation model's pretraining corpus and retain their original DFT reference labels~\cite{mace_mp0}.
Pseudolabel replay is an implementation of synthetic replay~\cite{li2016learning}, replacing the original replay reference labels with predictions from the pretrained foundation model itself.
A major advantage of this approach is that it decouples the replay-data source from the choice of foundation model: any structurally diverse dataset can be used to generate pseudolabels with the foundation model that is being fine-tuned, even when its original training data are unavailable.
In addition, the use of pseudolabels may reduce the likelihood that the replay head overfits the replay data, which could happen when the replay-data subset is much less diverse than the full foundation model pretraining dataset.
By eliminating the original reference labels, the replay head is restricted to fitting the foundation model results, rather than continuing to refine the fit toward the reference data values. The empirical consequences of replay data choice are reported in the Discussion and in Appendix~\ref{app:replay_data_ablation}.

\subsection{Atomic reference energy initialisation}
\label{sec:e0_method}

The total energy of a chemical system consists mostly of core-electron contributions, which are large, composition-dependent terms that are essentially insensitive to atomic arrangements or bonds. Since these contributions carry no information about chemical interactions, referencing energies to those of the isolated atoms removes this inert baseline, leaving only the much smaller atomisation energy for the model to learn.
Therefore, MACE models predict these atomisation energies referenced to the energies of isolated atoms in vacuum rather than absolute total energies. Per-element atomic reference energies (E0s) are stored by the model and added to its output during training and inference; the model therefore learns to predict interaction energies relative to the non-interacting baseline. Although isolated atoms represent a chemical extreme far removed from the condensed-phase configurations that dominate training sets, this convention substantially improves model stability and enables transfer across levels of theory. When changing the level of theory between pretraining and fine-tuning, the dominant source of label inconsistency is the shift in these reference energies; when fine-tuning on data computed with a different DFT functional, code, pseudopotential, or basis set, manually re-referencing the E0s therefore absorbs most of the systematic offset, leaving the remaining interaction energies as good initial guesses for the fine-tuning procedure. Notably, the non-linear architecture introduced in MACE-MH1 includes learnable bias terms that contribute to isolated-atom energies alongside the explicit E0 parameters, so the isolated-atom reference is not fully fixed by the E0s alone.
The essential requirement for the fine-tuning fit is that the E0s have the same meaning as those of the underlying foundation model, which, for all currently recommended ones, are those of spin-polarised \emph{isolated} atoms.

\paragraph{Explicit isolated-atom calculations.} The E0s are computed directly by performing single-atom DFT calculations using the same code and level of theory as the fine-tuning data, with spin polarisation in a simulation cell without any imposed symmetry and large enough that periodic images do not interact. These unambiguously define the reference state used during foundation model pretraining, but require access to the original DFT workflow, which is not always available---for example, when fine-tuning on a published community dataset.

\paragraph{Averaging-based E0 estimation.} One approach to estimating E0s is to solve a linear least-squares system that fits the fine-tuning energies as a linear combination of per-element contributions:
\begin{equation}
\{E_0^Z\}_{\text{avg}} =
\lim_{\lambda \to 0^+}
\arg\min_{\{E_0^Z\}}
\left[
\sum_{i=1}^{N_{\text{ft}}} \left( E_i^{\text{ft}} - \sum_{Z} n_i^Z E_0^Z \right)^2
+ \lambda \sum_Z \left(E_0^Z\right)^2
\right],
\end{equation}
where the sum runs over fine-tuning configurations, $n_i^Z$ is the count of atoms of element $Z$ in configuration $i$, and $E_i^{\text{ft}}$ is the fine-tuning reference energy. This assigns each element the average per-atom contribution to the energies of \emph{interacting} configurations rather than \emph{isolated} atom energies, and assumes that energy contributions are linearly separable by atom type. When the composition matrix is rank-deficient---typically because all training configurations share the same atomic ratio---the per-element corrections are not uniquely determined and only the total energy offset is identifiable. The $\lambda \to 0^+$ term above represents the SVD least-squares convention used in practice to select the minimum-norm solution when the system is rank-deficient.

\paragraph{Model-aware reestimation.} The reestimation approach~\cite{mh-1} leverages the pretrained model's predictions to find per-element E0 corrections that align the pretrained energy zero with the fine-tuning data. Given a pretrained model with original E0 values $\{E_0^Z\}_{\text{pre}}$, we seek corrections $\{\Delta E_0^Z\}$ that minimise the residual between the pretrained model's predictions and the fine-tuning reference energies:
\begin{equation}
\{\Delta E_0^Z\} =
\lim_{\lambda \to 0^+}
\arg\min_{\{\Delta E_0^Z\}}
\left[
\sum_{i=1}^{N_{\text{ft}}} \left( E_i^{\text{pre}} - \sum_{Z} n_i^Z \Delta E_0^Z - E_i^{\text{ft}} \right)^2
+ \lambda \sum_Z \left(\Delta E_0^Z\right)^2
\right],
\end{equation}
where $E_i^{\text{pre}}$ is the pretrained model's energy prediction for fine-tuning configuration $i$ and $E_i^{\text{ft}}$ is the corresponding fine-tuning reference energy. This is a linear least-squares problem yielding updated reference energies $E_0^{Z, \text{new}} = E_0^{Z, \text{pre}} + \Delta E_0^Z$. The procedure is meaningful only because the pretrained model already gives a roughly accurate estimate of the interaction energy; for a randomly initialised model, the residual would contain arbitrary model error rather than a chemically interpretable per-element offset. In the limiting case of a perfect pretrained model that predicts all interaction energies exactly, the residual is purely the difference between reference atomic baselines, and reestimation recovers the target DFT E0s exactly. This remains meaningful even when the target dataset uses all-electron E0 values, for which the absolute atomic baselines can differ greatly from those used during pretraining, because the fit separates that baseline shift from the pretrained interaction-energy prediction. Reestimation shares the same degeneracy as the averaging approach: when the system is rank-deficient, in this case we select the minimum-norm solution for the E0 corrections.

\section{Results}

We begin by evaluating key prerequisites for stable and effective fine-tuning: foundation model selection (Section~\ref{sec:foundation}), E0 initialisation (Section~\ref{sec:e0}), and method-specific hyperparameters (Section~\ref{sec:hypers}). These factors are examined first, as they establish the conditions under which meaningful comparisons between fine-tuning methods can be made. 

In what follows we use \emph{training setup} to refer to the full set of training choices that surround---but do not include---the choice of fine-tuning algorithm itself: foundation model selection, atomic reference energy (E0) initialisation, optimiser and its hyperparameters (learning rate, EMA decay, gradient clipping), loss-function weights, and any data normalisation or regularisation conventions. These prerequisites are treated as independent considerations rather than ordered in importance. Hyperparameters for each method are individually optimised to achieve stable and well-converged training dynamics, with final values summarised in Table~\ref{tab:hyperparams} (Appendix~\ref{app:hypers}). Only after these prerequisites are met do we compare the fine-tuning methods themselves, evaluating target-task accuracy, out-of-distribution transfer, and forgetting across the benchmark systems (Section~\ref{sec:method_comparison}).

\subsection{Foundation model quality}
\label{sec:foundation}
\begin{figure}[h]
    \centering
    \begin{subfigure}[b]{0.62\textwidth}
        \centering
        \includegraphics[width=\linewidth]{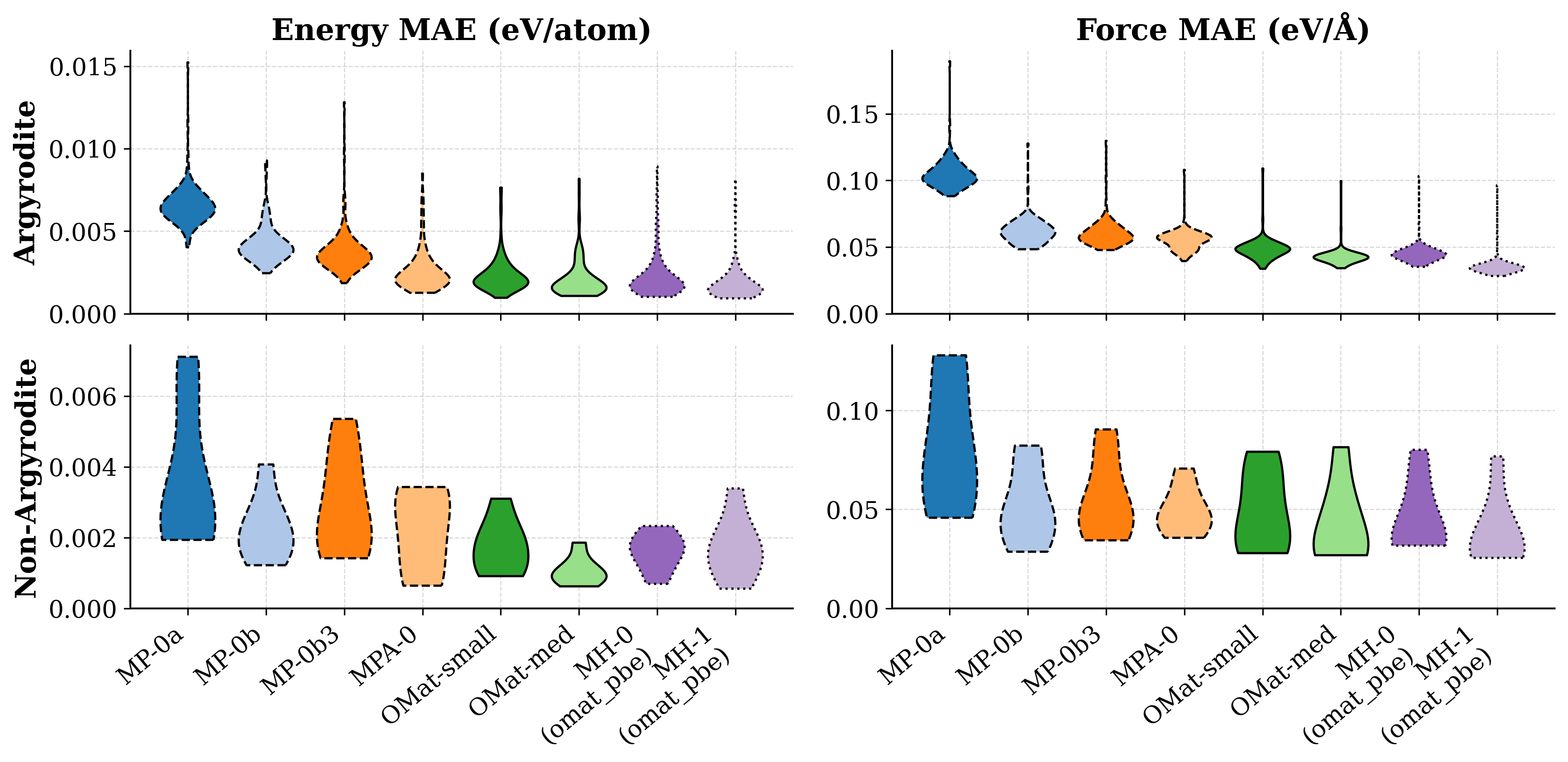}
        \caption{Lithium electrolyte.}
        \label{fig:foundation_model_quality_a}
    \end{subfigure}\hfill
    \begin{subfigure}[b]{0.36\textwidth}
        \centering
        \includegraphics[width=\linewidth]{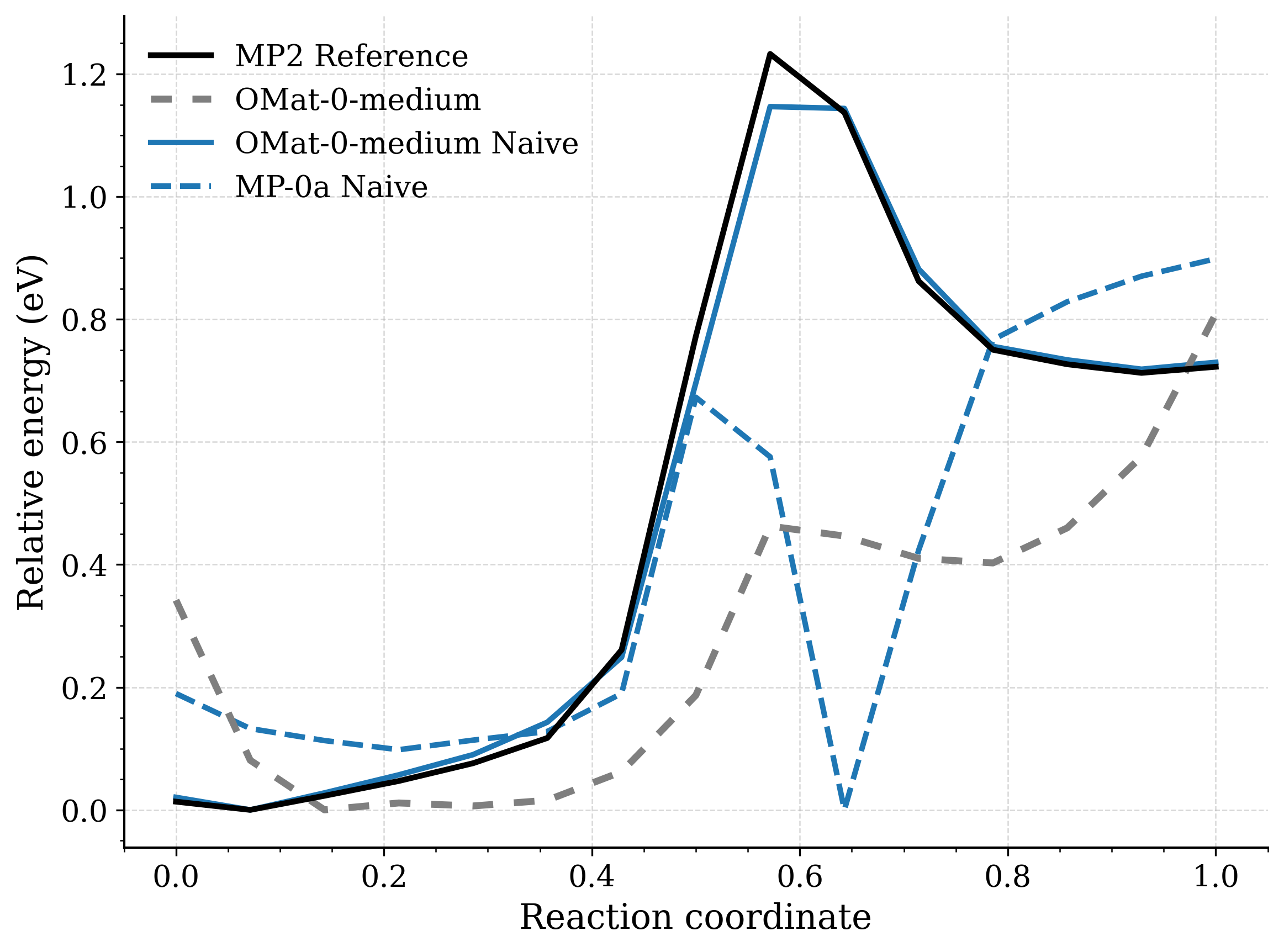}
        \caption{S$_\mathrm{N}$2 reaction.}
        \label{fig:foundation_model_quality_b}
    \end{subfigure}
    \caption{\textbf{Foundation model quality strongly shapes fine-tuning outcomes.} \textbf{(a) Lithium electrolytes:} Multihead-pseudolabel fine-tuning of several MACE foundation models on the lithium solid electrolyte system. Violin plots show force and energy errors on two held-out evaluation sets, both unseen during fine-tuning: other argyrodite compositions (top row) and non-argyrodite structures (bottom row). \textbf{(b) S$_\mathrm{N}$2 reaction:} reaction energy profiles comparing OMat24- and MPTraj-based foundation models. The OMat24-based naive fine-tuned model accurately reproduces the MP2 reference barrier, while the MPTraj-based model produces an unstable profile under identical NEB settings.}
        \label{fig:foundation_model_quality}
\end{figure}

We first examine the extent to which fine-tuning performance is set by the pretrained foundation itself. This comparison uses two benchmark tasks introduced in Section~\ref{sec:benchmarks}: the lithium electrolyte system, where models are fine-tuned on 500 LPSC configurations and evaluated on held-out lithium electrolytes, and the S$_\mathrm{N}$2 reaction, where reactant/product fine-tuning is assessed by NEB barrier profiles against MP2 reference data. 

The goal of this section is to isolate the effect of the foundation checkpoint with the fine-tuning recipe held fixed; method-level comparisons are deferred to Section~\ref{sec:method_comparison}.

For the lithium electrolyte benchmark (Fig.~\ref{fig:foundation_model_quality}a), all models are fine-tuned using the same pseudolabel replay protocol and the same 10,000 random configurations from MPTraj, so the comparison isolates the effect of the foundation model as much as possible. The held-out argyrodite set provides the clearest trend. Although these structures share the argyrodite structure with LPSC, they differ in composition. Across both energy and force MAE, the newer foundations give systematically lower errors than the earliest MP0a model: MP0b, MP0b3, and MPA-0 already reduce the error substantially, and the OMat- and MH-family models are among the best-performing models. Thus, even when the fine-tuning dataset and algorithm are fixed, improved foundation models translate directly into better out-of-distribution transfer across related electrolyte chemistries.

The non-argyrodite evaluation set is a more difficult test because both composition and crystal structure move further from the LPSC fine-tuning distribution. Here, the ordering shares the same broad pattern: not every newer model improves every metric on every non-argyrodite structure, but better and broader foundations improve the typical extrapolative behaviour. 

For S$_\mathrm{N}$2 reaction barriers (Fig.~\ref{fig:foundation_model_quality}b), the fine-tuned OMat24-based model accurately reproduces the MP2 reference barrier height, while the fine-tuned MPTraj-based model produces erratic and inaccurate reaction profiles under identical nudged elastic band parameters. Notably, OMat24 contains no molecular systems in its pretraining corpus, whereas MPTraj contains a small number, so this advantage cannot be attributed to prior exposure to chemically similar structures; instead, it demonstrates that richer coverage of inorganic chemical space during pretraining transfers to improved molecular fine-tuning outcomes---consistent with the finding that foundation models trained on inorganic crystals extrapolate broadly across chemistry~\cite{mace_mp0}. 

Taken together, these results argue that foundation-model quality should be treated as a primary design choice in fine-tuning workflows rather than a fixed background detail. The gains seen here are consistent with improvements in pretraining data scale, diversity, label consistency, and architecture being inherited by downstream fine-tuned models. We therefore expect this pattern to persist as foundation models improve: for applications that require transfer outside a narrow training distribution, investing in the strongest available foundation is likely to remain at least as important as choosing among specialised fine-tuning algorithms.

\subsection{Atomic reference energy initialisation}
\label{sec:e0}

\begin{figure}[h]
    \centering
    \begin{subfigure}[b]{0.55\textwidth}
        \centering
        \includegraphics[width=\linewidth]{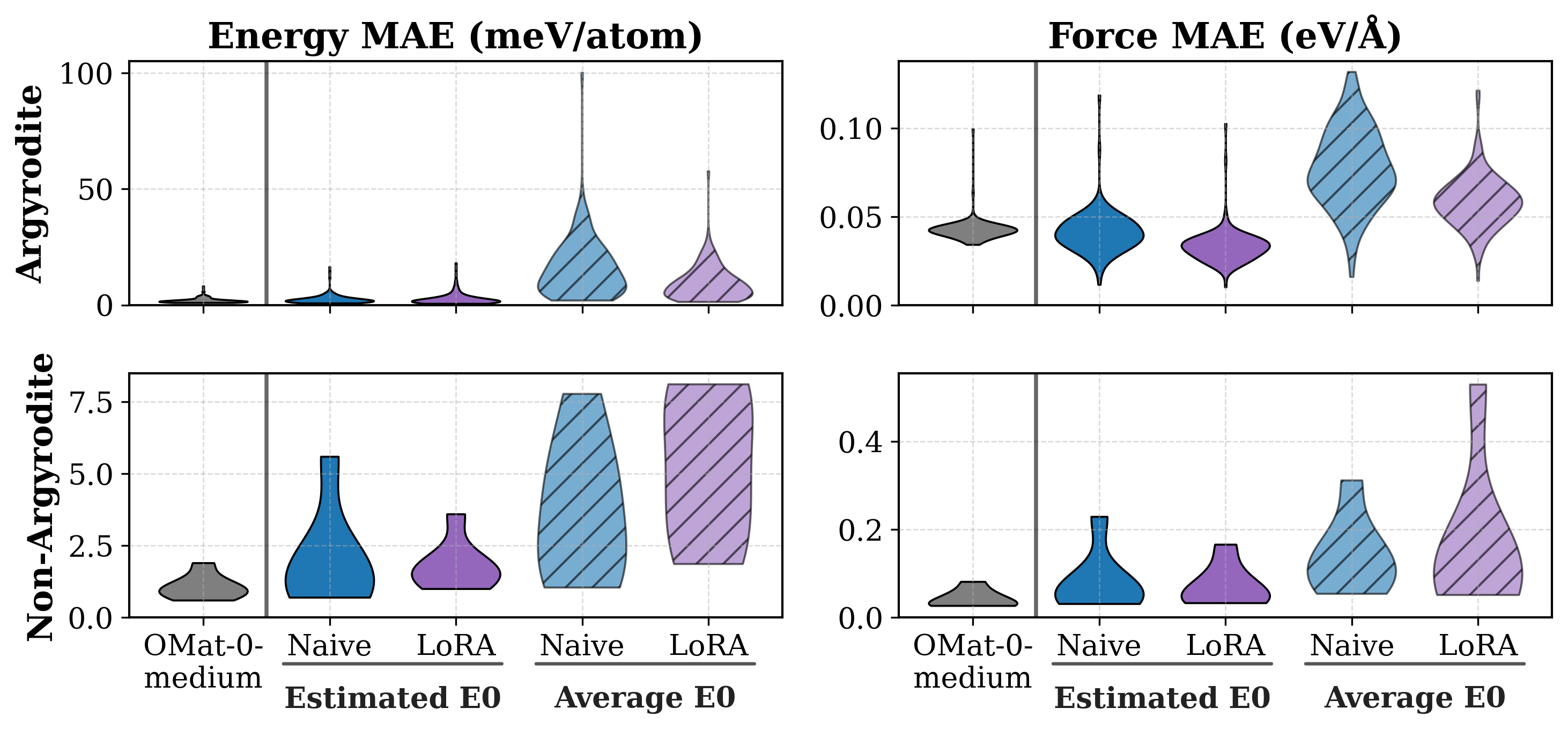}
        \caption{Lithium electrolyte.}
        \label{fig:e0_combined_a}
    \end{subfigure}
    \begin{subfigure}[b]{0.4\textwidth}
        \centering
        \includegraphics[width=\linewidth]{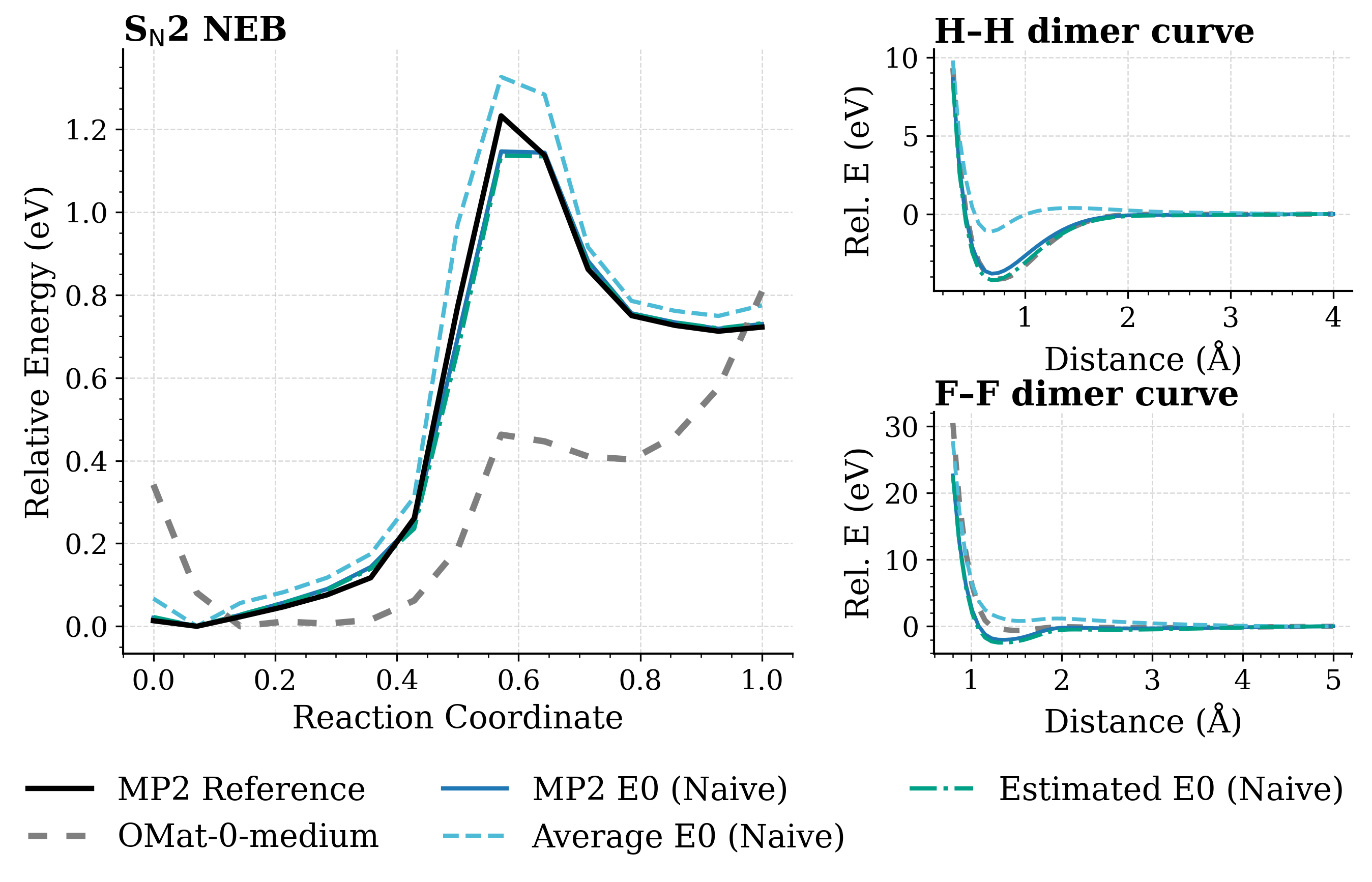}
        \caption{S$_\mathrm{N}$2 reaction.}
        \label{fig:e0_combined_b}
    \end{subfigure}
    \caption{\textbf{E0 initialisation can exceed method-level differences.} \textbf{(a) Lithium electrolyte (LPSC):} reestimated versus averaged E0s for Naive and LoRA fine-tuning with the MACE-OMat-0-medium foundation. Models trained with averaged E0s exhibit substantially higher errors and wider distributions on both held-out test sets (other argyrodites and non-argyrodites, both unseen during fine-tuning). \textbf{(b) S$_\mathrm{N}$2 reaction:} NEB and dimer energy curves comparing E0 choices. Reestimated and explicit isolated-atom E0s yield nearly identical curves, whereas averaged E0s distort the interaction profile.}
        \label{fig:e0_combined}
\end{figure}

Atomic reference energy initialisation is one of the most consequential prerequisites for stable fine-tuning. The three procedures we compare---explicit isolated-atom calculations, averaging-based estimation, and our model-aware reestimation implementation---are described in Section~\ref{sec:e0_method}. Here we evaluate their empirical consequences.

Figure~\ref{fig:e0_combined} demonstrates that E0 choice has a larger impact than the differences between fine-tuning methods. For the lithium electrolyte system with MACE-OMat-0-medium, models trained with averaged E0s achieve force RMSEs 2--3$\times$ higher than models trained with reestimated E0s, regardless of whether naive or LoRA fine-tuning is used. The spread in performance across different fine-tuning methods with \emph{correct} (i.e. explicitly calculated or reestimated) \emph{E0s} is smaller than the degradation caused by using averaged E0s with \emph{any} method. Reestimated E0s consistently yield lower errors on both the other-argyrodite and the non-argyrodite test sets---both of which are out of distribution relative to the LPSC training data---demonstrating that the benefit is not overfitting to the training distribution but rather providing a coherent energy reference that enables the model to learn transferable representations.

\begin{figure}[h]
    \centering
    \includegraphics[width=0.8\textwidth]{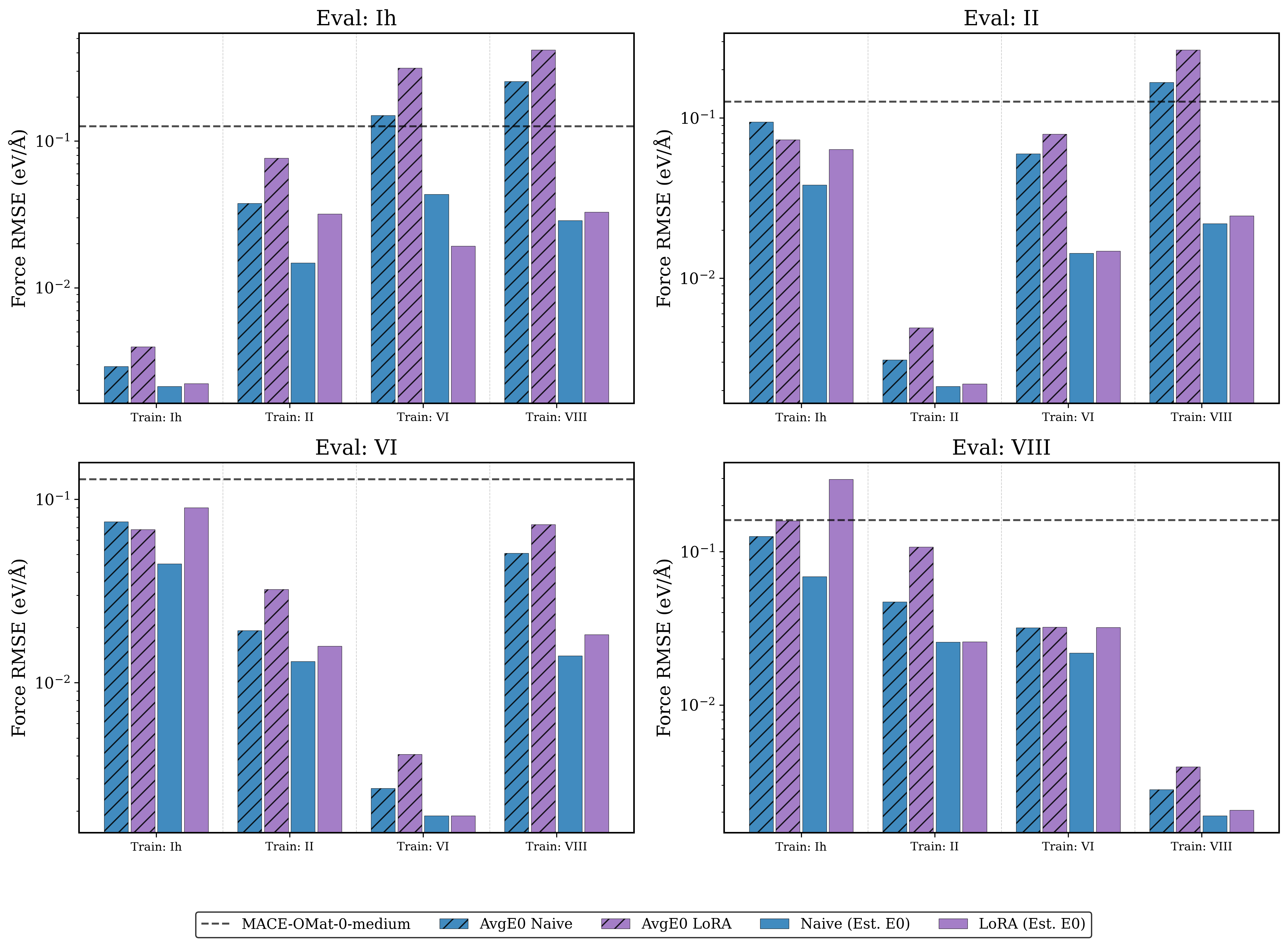}
    \caption{Impact of E0 initialisation on ice cross-phase learning with MACE-OMat-0-medium. Force RMSE for models trained on each ice phase (Ih, II, VI, VIII) and evaluated on all phases, comparing reestimated E0s versus averaged E0s for Naive and LoRA fine-tuning. Reestimated E0s consistently yield lower errors across all phase combinations, with the improvement exceeding the differences between fine-tuning methods.}
        \label{fig:e0_comparison_water}   
\end{figure}

The ice cross-phase system (Fig.~\ref{fig:e0_comparison_water}) follows this pattern: across all four ice phases and both training conditions (Naive and LoRA), reestimated E0s consistently produce lower force RMSEs than averaged E0s, with improvements larger than the differences between fine-tuning methods in this comparison.
More importantly, incorrect E0 initialisation is sufficient to make ice models that appear successful based on validation RMSE fail in molecular dynamics (Table~\ref{tab:e0_stability}). For ice systems trained with averaged E0s, MD trajectories at 100~K frequently become unstable within 50~ps regardless of fine-tuning method, with structures developing unphysical distortions or bond breaking. In contrast, with reestimated E0s, all tested fine-tuning methods (naive, LoRA, multihead) produce stable trajectories extending beyond 250~ps at 100~K and remain stable during temperature ramps to 800~K.

\begin{table}[h]
\centering
\begin{tabular}{lcccc}
\toprule
\multirow{2}{*}{\textbf{Method}} & \multicolumn{2}{c}{\textbf{Averaged E0s}} & \multicolumn{2}{c}{\textbf{Reestimated E0s}} \\
\cmidrule(lr){2-3} \cmidrule(lr){4-5}
 & \textbf{100~K Stable} & \textbf{T-ramp Stable} & \textbf{100~K Stable} & \textbf{T-ramp Stable} \\
\midrule
From-scratch & \cmark & \xmark & n/a & n/a \\
Naive & \xmark & \xmark & \cmark & \cmark \\
LoRA ($r{=}4$) & \xmark & \xmark & \cmark & \cmark \\
Multihead & \xmark & \xmark & \cmark & \cmark \\
\bottomrule
\end{tabular}
\caption{\textbf{E0 initialisation determines MD stability.} Stability of ice Ih models trained on 25 configurations with averaged versus reestimated E0s. ``100~K Stable'': NPT simulations at 100~K complete 250~ps without unphysical distortions. ``T-ramp Stable'': stability during temperature ramps from 50~K to 800~K over 50~ps. With averaged E0s, all fine-tuned models fail; with reestimated E0s, all succeed. \cmark~=~stable, \xmark~=~unstable.}
\label{tab:e0_stability}
\end{table}

\subsubsection{SPICE E0 reestimation}

To test whether model-aware E0 reestimation remains useful when the magnitudes of the atomic baseline energies are much larger than the interaction energies of interest, we applied the E0 averaging and E0 reestimation procedures to the full SPICE dataset, which uses all-electron energies, using the pretrained MACE-OMat-0-medium model, which was pretrained using valence-electron-only atom energies shifted to approximately zero as E0s. Note that we
apply the two sets of E0 values without fine-tuning or otherwise training the model on SPICE. Table~\ref{tab:spice_e0_comparison} compares the resulting averaged and reestimated E0s with explicit SPICE DFT isolated-atom reference energies for the elements present in our SPICE data.

\begin{table}[hbt!]
\centering
\small
\setlength{\tabcolsep}{4pt}
\begin{tabular}{lrrrrr}
\toprule
& \textbf{Explicit eval.} & \multicolumn{2}{c}{\textbf{Averaging}} & \multicolumn{2}{c}{\textbf{Reestimation}} \\
\cmidrule(lr){2-2} \cmidrule(lr){3-4} \cmidrule(lr){5-6}
\textbf{Element} & \textbf{E0 (eV)} & \textbf{E0 (eV)} & \textbf{$\Delta$ (eV)} & \textbf{E0 (eV)} & \textbf{$\Delta$ (eV)} \\
\midrule
H (1)   & -13.572    & -16.385    & -2.813 & -13.753    & \textbf{-0.181} \\
C (6)   & -1030.567  & -1037.582  & -7.015 & -1030.268  & \textbf{0.299}  \\
N (7)   & -1486.375  & -1490.736  & -4.361 & -1485.902  & \textbf{0.473}  \\
O (8)   & -2043.934  & -2048.170  & -4.237 & -2043.254  & \textbf{0.680}  \\
F (9)   & -2715.319  & -2718.470  & -3.152 & -2714.923  & \textbf{0.396}  \\
P (15)  & -9287.407  & -9291.485  & -4.078 & -9287.549  & \textbf{-0.141} \\
S (16)  & -10834.485 & -10836.893 & -2.409 & -10834.133 & \textbf{0.351}  \\
Cl (17) & -12522.649 & -12524.330 & -1.681 & -12522.487 & \textbf{0.163}  \\
Br (35) & -70045.284 & -70046.423 & -1.139 & -70045.046 & \textbf{0.238}  \\
I (53)  & -8102.525  & -8102.956  & -0.431 & -8102.156  & \textbf{0.369}  \\
\bottomrule
\end{tabular}
\caption{\textbf{SPICE atomic reference energies from averaging and model-aware reestimation.} Values are obtained by applying E0 averaging and model-aware reestimation to the full SPICE dataset using the pretrained MACE-OMat-0-medium model, without training on SPICE. The table includes only elements present in our SPICE data. Deltas are computed relative to the explicit SPICE DFT isolated-atom E0s.}
\label{tab:spice_e0_comparison}
\end{table}

Despite absolute atomic reference energies ranging from tens to tens of thousands of eV, reestimation recovers values much closer to the explicit SPICE DFT E0s than averaging. Across the elements in Table~\ref{tab:spice_e0_comparison}, the mean absolute E0 error is reduced from 3.13~eV for averaging to 0.33~eV for reestimation. This shows that the procedure is not limited to datasets whose total energies have atomic baselines comparable to the fine-tuned foundation model. It remains applicable when large per-element atomic contributions dominate the absolute energy scale, because it fits the model-consistent offset needed to align the pretrained interaction model with the target reference. The result is notable because the starting MACE-OMat-0-medium foundation was pretrained on inorganic crystals, whereas SPICE contains organic and biomolecular configurations with substantially different atomic environments.

\subsubsection{Models with learnable E0s}

For some applications, it is important not only to obtain accurate relative energetics, but also to recover accurate isolated-atom energies. An example is the computation of defect formation energies, which depends explicitly on atomic reference energies: a mismatch between the model's predicted isolated-atom energies and the true values at the target level of theory can introduce large systematic errors, even when total energies are otherwise well reproduced. 
The MACE-MH1 model uses a modified architecture and is a special case for E0 handling: it includes learnable bias terms that are summed with the explicit E0s to determine the isolated-atom energies. As a result, the isolated-atom reference is shared between the explicit E0s and these biases, and accurate isolated-atom predictions are not guaranteed by setting the E0s correctly alone.
The E0 reestimation procedure remains applicable in this setting, since it still aims to recover the best estimate of the isolated-atom energies given the pretrained model and the available training data. However, its role is primarily to improve training dynamics. Initialising the explicit E0s with reestimated values close to the target level of theory gives the model a more appropriate starting point, leading to better-conditioned training. However, because the learnable biases can still absorb part of the isolated-atom reference during training, we recommend also including true target-level isolated-atom energies as explicit training points whenever accurate atomic energies matter for the downstream application, even when isolated-atom DFT energies are available and used to initialise the E0s.

\begin{table}[hbt!]
\centering
\begin{tabular}{l c c c}
\toprule
& \textbf{Explicit eval.} & \textbf{Averaging} & \textbf{Reestimation} \\
\cmidrule(lr){2-2} \cmidrule(lr){3-3} \cmidrule(lr){4-4}
\textbf{Repeats} & \textbf{RMSE E0 (meV)} & \textbf{RMSE E0 (meV)} & \textbf{RMSE E0 (meV)} \\
\midrule
0  & \textbf{15.0} & 3855.1 & 252.7        \\
1  & \textbf{11.8} & 47.8   & 31.2         \\
5  & \textbf{4.7}  & 16.8   & 5.6          \\
10 & \textbf{1.1}  & 17.5   & 2.7          \\
20 & \textbf{1.0}  & 9.2    & 2.7          \\
40 & 1.8           & 13.3   & \textbf{0.6} \\
\bottomrule
\end{tabular}
\caption{\textbf{Isolated-atom RMSE summary for MACE-MH1 multihead fine-tuning across E0 choices.} All rows are multihead replay fine-tuning of MACE-MH1 with DFT replay labels. Columns differ only in the E0 convention used for the target task (explicitly evaluated with DFT, training-set average, or reestimated). Rows show the number of repeated isolated-atom configurations included in the fine-tuning data. Values are in meV/atom; lower is better.}
\label{tab:universal_data}
\end{table}

Table~\ref{tab:universal_data} reports the resulting isolated-atom RMSE for explicitly evaluated, averaged, and reestimated E0s, as a function of the number of repeated isolated-atom configurations included in the training data. All rows are multihead replay runs of MACE-MH1 with DFT replay labels. The target dataset combines DFT isolated-atom energies with UiO-66 metal--organic framework configurations from Ref.~\cite{Elena2025}, augmented with our own data; additional loss-function variants are reported in Appendix~\ref{app:mh1_stress} (Table~\ref{tab:stress_data}).
The 10-repetition setting was also tested with longer training. Extending the MACE-MH1 multihead run with DFT replay labels to 20 epochs reduced the isolated-atom RMSE for reestimated E0s from 2.7 to 1.0~meV/atom, while the corresponding error using DFT-evaluated E0s increased from 1.1 to 1.3~meV/atom. Thus, with sufficient isolated-atom exposure and training, reestimated E0s reached isolated-atom accuracy comparable to explicit DFT-evaluated E0s in this system.
Together, these results make E0 consistency a prerequisite rather than a minor preprocessing detail, while the MACE-MH1 case shows that models with learnable isolated-atom contributions may also require explicit isolated-atom training configurations.

\subsection{Hyperparameters}
\label{sec:hypers}

Each method's hyperparameter configuration was individually optimised to ensure stable training dynamics and well-converged models, with the complete settings reported in Table~\ref{tab:hyperparams}. Hyperparameters were tuned separately for each method, and only the best-performing configurations are presented, since adopting a single shared setting across methods could confound performance comparisons by introducing effects attributable to hyperparameter misconfiguration rather than the methods themselves. Several noteworthy patterns emerged from this optimisation process, which are discussed in detail in this section.

\textbf{Weight decay.} For fine-tuning, weight decay should be set to zero, since it penalises the magnitude of the parameters and therefore pulls them toward zero rather than toward the pretrained values. This means that weight decay actively drives the model \emph{away} from the foundation model's learned solution, counteracting one of the main benefits of starting from a pretrained checkpoint.

\textbf{Learning rate.} The optimal learning rate varies substantially across methods (Table~\ref{tab:hyperparams}). Empirically, LoRA tolerates learning rates an order of magnitude higher than naive fine-tuning ($10^{-2}$ vs $10^{-3}$); this mirrors findings in other domains where constrained adaptation permits more aggressive optimisation~\cite{thinky}.
Multihead replay training, by contrast, requires a substantially lower learning rate ($\sim$$10^{-4}$); higher learning rates lead to worse convergence and worse final accuracy on both the target and replay objectives, perhaps because the model receives competing update signals from the two heads at each step. A learning-rate sweep documenting this behaviour is reported in Appendix~\ref{app:mh_lr_ablation} (Fig.~\ref{fig:mh_lr_ablation}).

\textbf{Loss function weights.} Two-stage loss scheduling---first emphasising force matching, then switching to energy matching---is common practice in from-scratch MLIP training~\cite{medicineMACE}, where forces provide dense per-atom gradient signals that help a randomly initialised model establish a reasonable local PES before global energy fitting refines it. For fine-tuning, where the model already starts with accurate forces and a coherent PES, this staged approach can destabilise training at the transition point (an example is shown in Appendix Fig.~\ref{fig:loss_schedule_ablation}). Additionally, we did not observe a benefit of two-stage scheduling compared to single-stage training with constant weights.

For our fine-tuning experiments, we use $\lambda_E = 10, \lambda_F = 10$ on the target task throughout training, placing higher weight on energy accuracy, which is important for downstream applications such as barrier calculations and relative phase stabilities. For multihead replay training, we apply $\lambda_E^{\text{replay}} = 1, \lambda_F^{\text{replay}} = 10$ on the replay head, matching the force-focused weighting used during the foundation model's original pretraining. These weights remain constant throughout training.
A loss-weight schedule comparison supporting this recommendation, comparing constant target weights against two-stage schedules under otherwise matched fine-tuning settings, is reported in Appendix~\ref{app:loss_schedule_ablation}.

\subsection{Fine-tuning method comparison by objective}
\label{sec:method_comparison}

With prerequisites established, MACE-OMat-0-medium is selected as the primary foundation model and reestimated E0s are adopted as the reference energy convention for all subsequent experiments, where explicit isolated-atom E0s are unavailable. Now we compare fine-tuning methods by what the fine-tuned model is meant to achieve. We separate four objectives:
\begin{enumerate}
    \item \textbf{In-distribution specialisation on a single system.} The fine-tuned model targets a single system, and out-of-distribution accuracy is irrelevant.
    \item \textbf{Near-transfer within a related chemical family.} A broader objective requiring the model to generalise across compositions and structures not seen during fine-tuning but related to the training data.
    \item \textbf{Transfer to a different chemistry.} Adaptation to chemical regions distinct from the foundation-model pretraining data, especially relative to from-scratch training when more data are available.
    \item \textbf{Preservation of foundation-model behaviour} away from the fine-tuning distribution.
\end{enumerate}

\textbf{1. In-distribution specialisation on a single system.}

\begin{table*}[h]
\centering
\caption{\textbf{Validation RMSE for the two narrow-task systems.} S$_\mathrm{N}$2 results are on the full training dataset (single run); aqueous NaCl results are at 10\% training data, reported as mean $\pm$ std over seeds 1--3. Bold marks the lowest entry in each (system, column) pair.}
\label{tab:rmse_narrow}
\begin{tabular}{lcccc}
\toprule
& \multicolumn{2}{c}{\textbf{S$_\mathrm{N}$2}}
& \multicolumn{2}{c}{\textbf{Aqueous NaCl}} \\
\cmidrule(lr){2-3} \cmidrule(lr){4-5}
\textbf{Method}
& \textbf{Energy RMSE}
& \textbf{Force RMSE}
& \textbf{Energy RMSE}
& \textbf{Force RMSE} \\
& \textbf{(meV/atom)}
& \textbf{(meV/\AA)}
& \textbf{(meV/atom)}
& \textbf{(meV/\AA)} \\
\midrule
Scratch & 0.19 & 12.1 & 0.47 $\pm$ 0.06 & 42.44 $\pm$ 5.28 \\
Naive & 0.05 & \textbf{3.8} & 0.45 $\pm$ 0.11 & 43.79 $\pm$ 5.03 \\
Freeze (5) & -- & -- & 0.46 $\pm$ 0.16 & 44.16 $\pm$ 5.06 \\
Freeze (6) & -- & -- & 0.74 $\pm$ 0.23 & 52.57 $\pm$ 3.83 \\
LoRA & \textbf{0.02} & 4.1 & \textbf{0.30 $\pm$ 0.04} & \textbf{41.54 $\pm$ 4.76} \\
Multihead & 0.05 & 4.5 & 0.43 $\pm$ 0.11 & 42.00 $\pm$ 4.58 \\
Pseudolabel & 0.03 & 4.3 & 0.42 $\pm$ 0.10 & 41.85 $\pm$ 4.65 \\
PS+LoRA & 0.04 & 6.5 & 0.55 $\pm$ 0.16 & 44.22 $\pm$ 4.04 \\
\bottomrule
\end{tabular}
\end{table*}

\begin{figure}[h]
    \centering
    
    \includegraphics[width=0.9\textwidth]{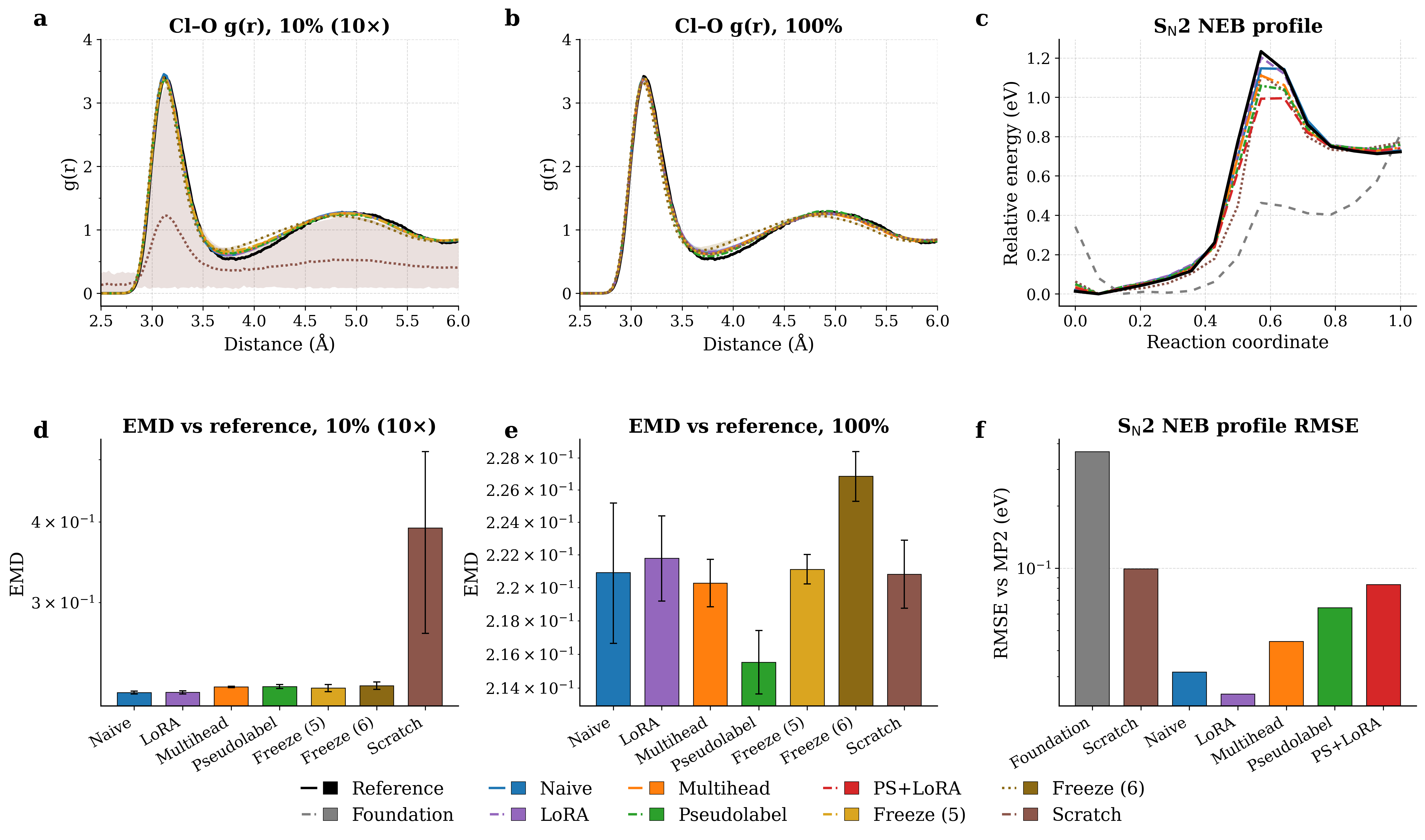}
    \caption{Narrow-task fine-tuning results. \textbf{a}, Cl--O RDFs at 10\% training data, averaged over three random seeds. \textbf{b}, Cl--O RDFs at 100\% training data. \textbf{c}, S$_\mathrm{N}$2 nudged elastic band energy profiles for models trained only on reactant/product configurations, excluding the transition state. Relative energy (eV) along the reaction coordinate is compared to the MP2 reference (dashed red). Naive and LoRA closely reproduce the reference barrier height and shape. Multihead and Pseudolabel slightly underestimate the barrier. From-scratch training substantially underestimates the barrier, while the foundation model before fine-tuning fails to capture the reaction profile. \textbf{d}, Dataset-size dependence of the earth mover's distance to the BPNN reference. All fine-tuning methods closely reproduce the reference solvation structure already at 10\% data, while training from scratch remains substantially worse.}
    \label{fig:nacl_rdf_10pct}
\end{figure}

We first test two single-system objectives from Section~\ref{sec:benchmarks}: S$_\mathrm{N}$2 reaction barriers and aqueous NaCl solvation structure.
Pointwise validation RMSE for both narrow-task systems is summarised in Table~\ref{tab:rmse_narrow}. At 10\% of the training data (Fig.~\ref{fig:nacl_rdf_10pct}), all fine-tuning methods closely reproduce the BPNN reference Cl--O solvation structure, with earth mover's distances (EMD) between 0.052 and 0.100, while training from scratch fails to capture the solvation shell structure (EMD = 0.685). Multihead achieves the lowest EMD among fine-tuning methods, and the Freeze variants show slightly elevated EMDs, suggesting that freezing embedding and message-passing layers limits adaptation of solvation-relevant features. The dataset-size dependence shows that Naive, LoRA, Multihead, and Pseudolabel converge to the reference RDF already at 10\% data, with minimal further improvement at 100\%; increasing training epochs at fixed dataset size yields only marginal gains. The full NaCl RDF and EMD panel is given in Appendix~\ref{app:nacl_full} (Fig.~\ref{fig:nacl_full_si}).

The RDFs show that narrow-task fine-tuning can perform very well on application-specific observables with limited data. Whether those same models remain physically sensible away from the evaluation manifold is addressed separately below through forgetting and RSS analyses.
\clearpage

\textbf{2. Transfer to related structures and compositions.}

\begin{figure}[h]
    \centering
    \begin{subfigure}[b]{0.53\textwidth}
        \centering
        \includegraphics[width=\linewidth]{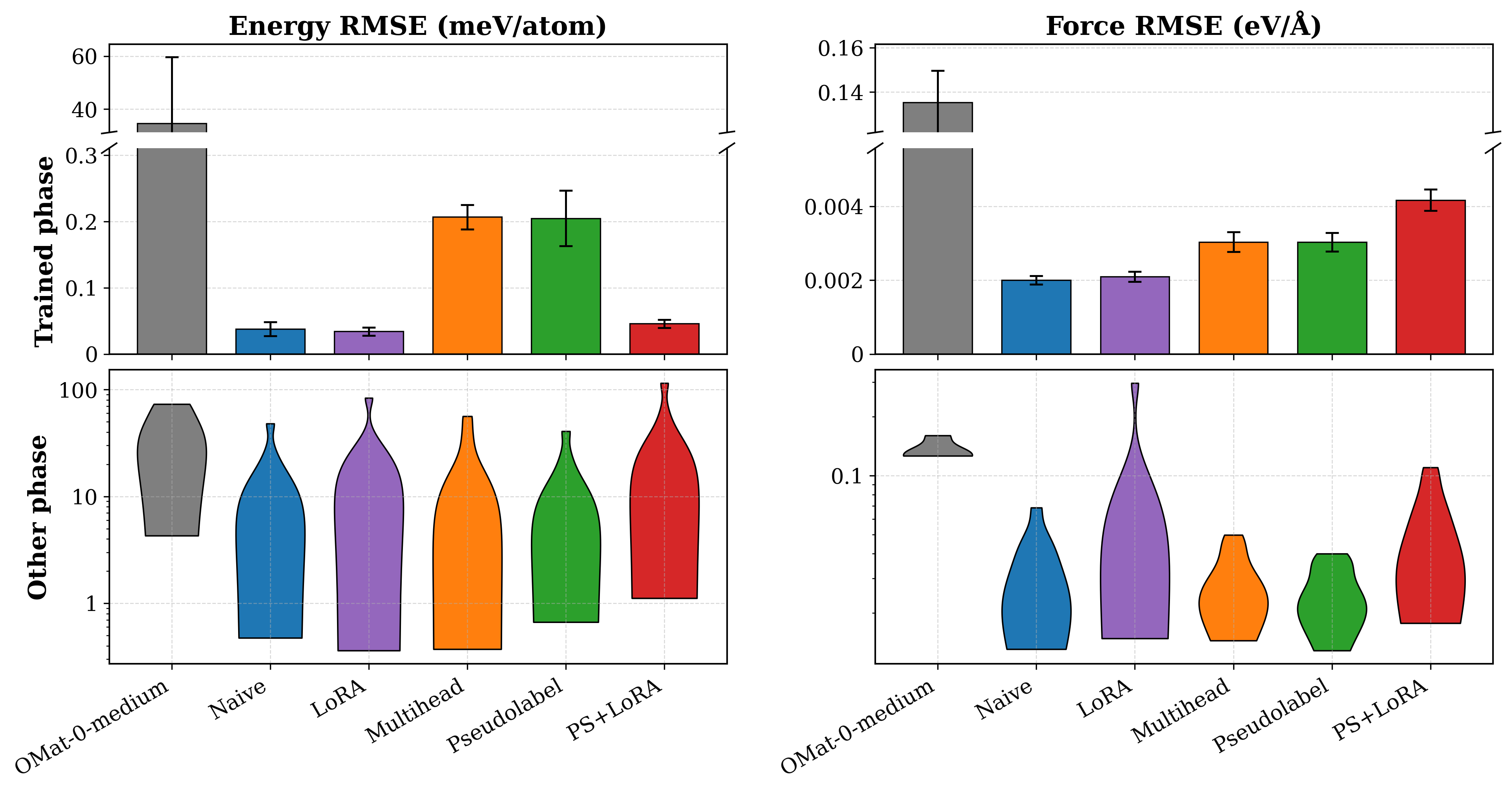}
        \caption{Ice polymorph cross-phase learning.}
        \label{fig:near_transfer_a}
    \end{subfigure}
    \begin{subfigure}[b]{0.4\textwidth}
        \centering
        \includegraphics[width=\linewidth]{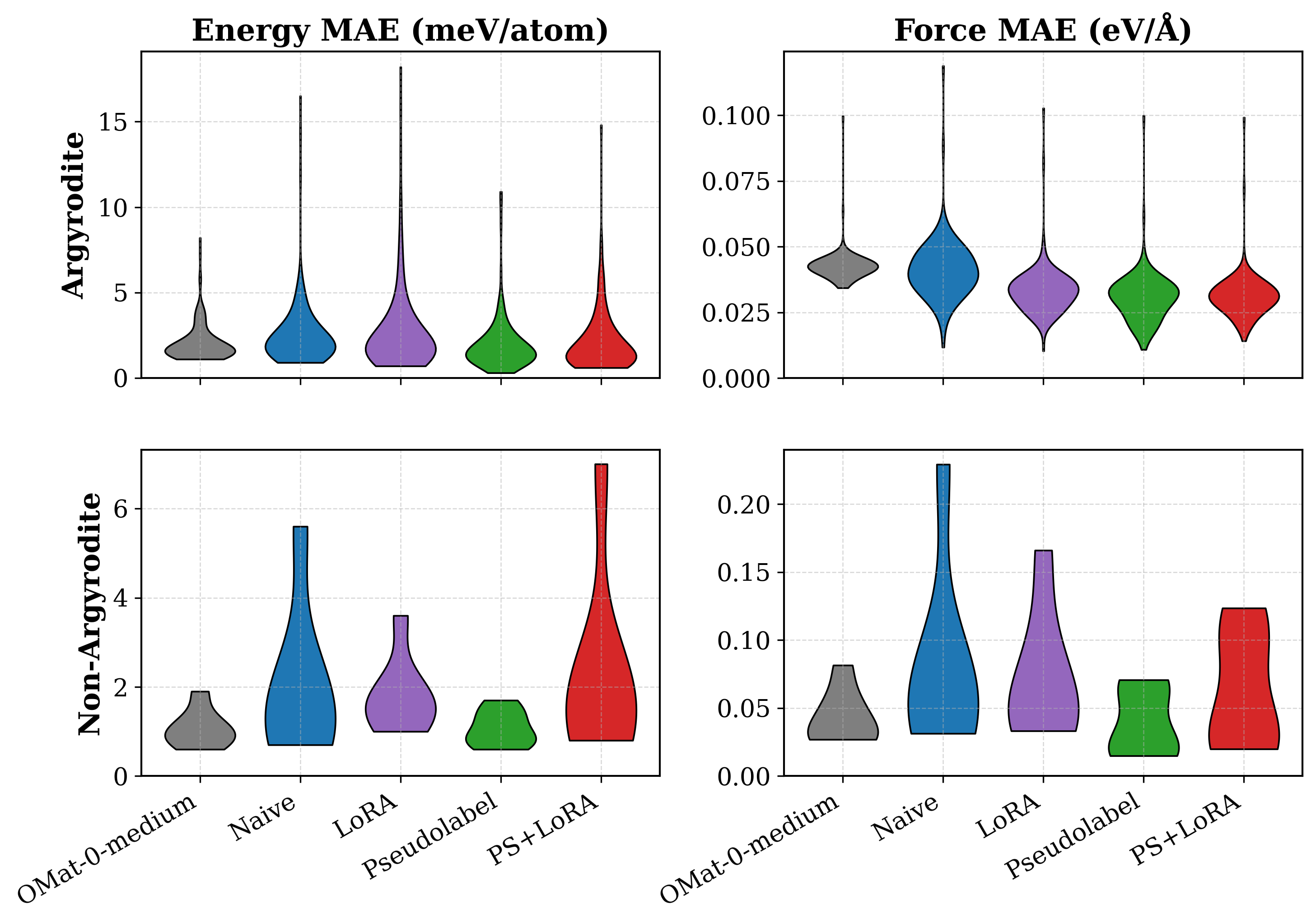}
        \caption{Lithium electrolyte.}
        \label{fig:near_transfer_b}
    \end{subfigure}
   \caption{Near-transfer within related chemical families. \textbf{(a) Ice polymorph cross-phase learning:} summarised as in-phase force RMSE and out-of-phase average force RMSE for each fine-tuning method with MACE-OMat-0-medium. \textbf{(b) Lithium electrolyte (LPSC) transfer:} with the MACE-OMat-0-medium foundation, comparing force MAE on two held-out test sets: other argyrodites and non-argyrodites, both unseen during fine-tuning.}
     \label{fig:near_transfer}
\end{figure}

We next consider transfer within related chemical families: cross-phase generalisation among ice polymorphs and transfer from LPSC to other lithium electrolytes. In ice (Fig.~\ref{fig:near_transfer}a), all methods achieve comparable in-phase force RMSEs, well below the foundation baseline, and the out-of-phase average differs only modestly between methods. In the lithium electrolyte system (Fig.~\ref{fig:near_transfer}b), all methods improve over the foundation baseline on the other-argyrodite test set, while replay-based training gives the best transfer to the non-argyrodite test set, the more distant of the two evaluation sets. Together these systems show that method-level differences are modest for near-distribution transfer but become more visible as the evaluation moves further from the fine-tuning distribution. The effect of foundation-model scale on ice cross-phase learning is shown separately in Appendix~\ref{app:ice_capacity} (Fig.~\ref{fig:water_cross_foundations_app}), and a single-phase versus combined-phase ice comparison is reported in Appendix~\ref{app:combined_vs_single} (Fig.~\ref{fig:combined_vs_single_si}).

\textbf{3. Fine-tuning across chemical domains.}

\begin{figure}[h]
    \centering
    \includegraphics[width=0.9\textwidth]{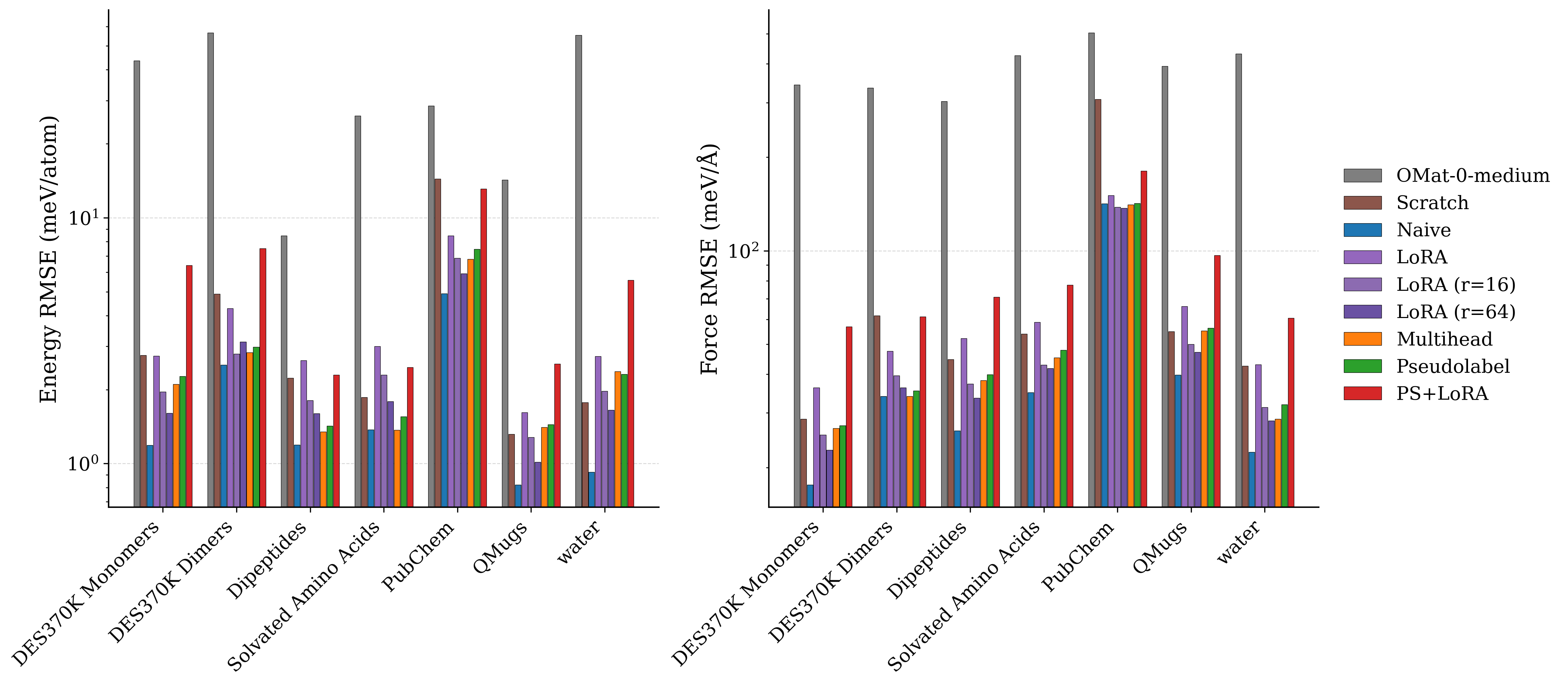}
   \caption{SPICE test set errors with 1 configuration per molecule training (19\,687 configurations total). Energy RMSE (left, meV/atom) and force RMSE (right, meV/\AA) are broken down by SPICE subset: DES370K monomers, DES370K dimers, dipeptides, solvated amino acids, PubChem, QMugs, and water.}
     \label{fig:spice_test_rmse}
\end{figure}

Fine-tuning on the SPICE dataset (Section~\ref{sec:benchmarks}) tests whether a foundation model trained predominantly on inorganic periodic structures can adapt to organic and biomolecular chemistry, and is the most demanding benchmark in our suite in terms of distance from the OMat24 pretraining domain.
On the SPICE test set itself in a low-data regime (1 configuration per unique molecule; 19\,687 configurations total, Figure~\ref{fig:spice_test_rmse}), all fine-tuning methods except PS+LoRA (pseudolabelled replay combined with LoRA) match or outperform from-scratch training across all subsets (DES370K monomers and dimers, dipeptides, solvated amino acids, PubChem, QMugs, and water clusters). The from-scratch baseline here is trained on the same SPICE subset as the fine-tuned models, so this comparison isolates the effect of pretrained initialisation rather than testing fine-tuning against a separately pretrained organic foundation model; with sufficient data and training time, from-scratch and naive fine-tuning are expected to converge to similar solutions. Naive fine-tuning achieves the best SPICE test accuracy, indicating that the pretrained model provides a useful starting point but must still adapt substantially to the organic and biomolecular domain. Higher-rank LoRA ($r=64$), Multihead, and Pseudolabel replay give similar intermediate performance, remaining better than from-scratch training but worse than Naive. Lower-rank LoRA and PS+LoRA perform worst, in several cases comparable to or worse than from scratch; the additional degradation of PS+LoRA relative to LoRA or replay alone, also observed in benchmarks such as the S$_\mathrm{N}$2 NEB profile, is consistent with over-constraining the fine-tuning update. The similarity in target-task accuracy across methods that differ markedly in forgetting behaviour underscores that forgetting and in-domain performance are largely decoupled: methods that preserve the pretrained distribution (Multihead, Pseudolabel) do so without sacrificing accuracy on the fine-tuning task.

\begin{figure}[h]
    \centering
    \includegraphics[width=0.95\textwidth]{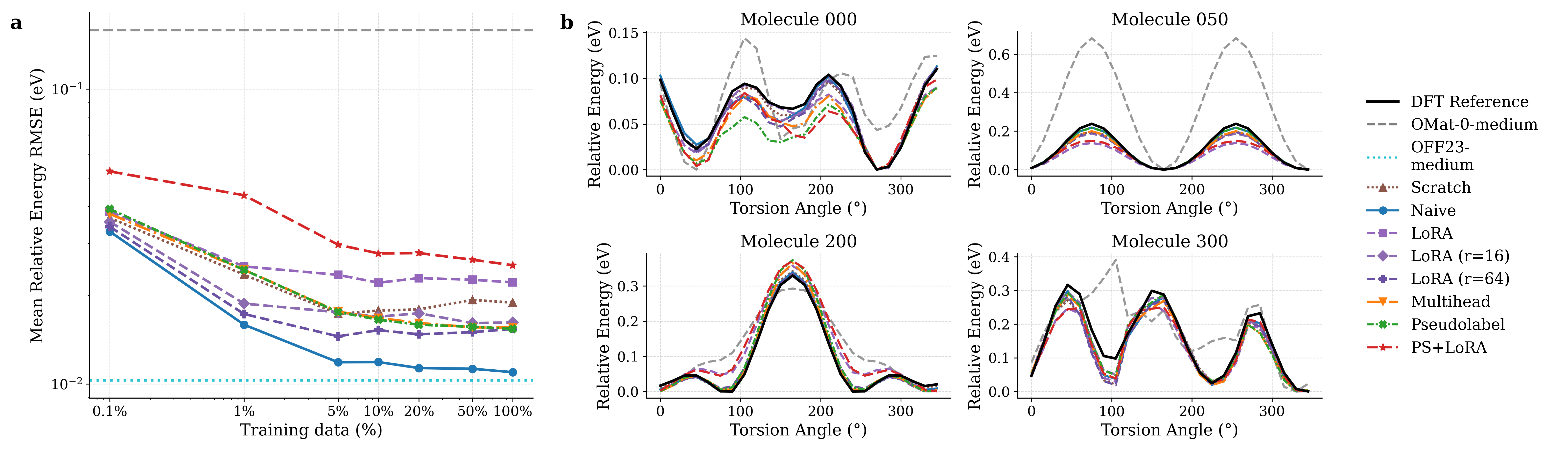}
   \caption{SPICE torsion accuracy across training methods. \textbf{a}, TorsionNet500 mean relative energy RMSE as a function of SPICE training data fraction. \textbf{b}, Representative torsion profiles comparing fine-tuned models, the foundation model before fine-tuning, and the DFT reference. Fine-tuned models substantially improve over the foundation model, and the best methods closely track the reference curves.}
     \label{fig:spice_torsion}
\end{figure}

Torsion profile accuracy probes one-dimensional slices of the PES along chemically meaningful degrees of freedom, providing a more structured test than individual energy and force predictions on held-out configurations. Figure~\ref{fig:spice_torsion} combines the data-quantity-scaling trend with representative torsion profiles.
Fine-tuned models substantially improve over the foundation model in reproducing DFT torsion energy profiles, and torsion accuracy improves monotonically with training data fraction for all methods before saturating. From-scratch and low-rank LoRA models exhibit the largest variability at small dataset sizes, consistent with insufficient capacity or data to capture the smooth one-dimensional energy variation along the dihedral coordinate evaluated on the fixed TorsionNet500 set. With increasing LoRA rank, the torsion profile accuracy improves.

\textbf{4. Foundation model quality preservation.}
The final objective is not downstream accuracy itself, but preserving desirable foundation-model behaviour after fine-tuning. This includes retaining accuracy on the original pretrained distribution and maintaining a physically sensible repulsive wall away from the fine-tuning data manifold.

\begin{figure}[t]
    \centering
    \includegraphics[width=0.95\textwidth]{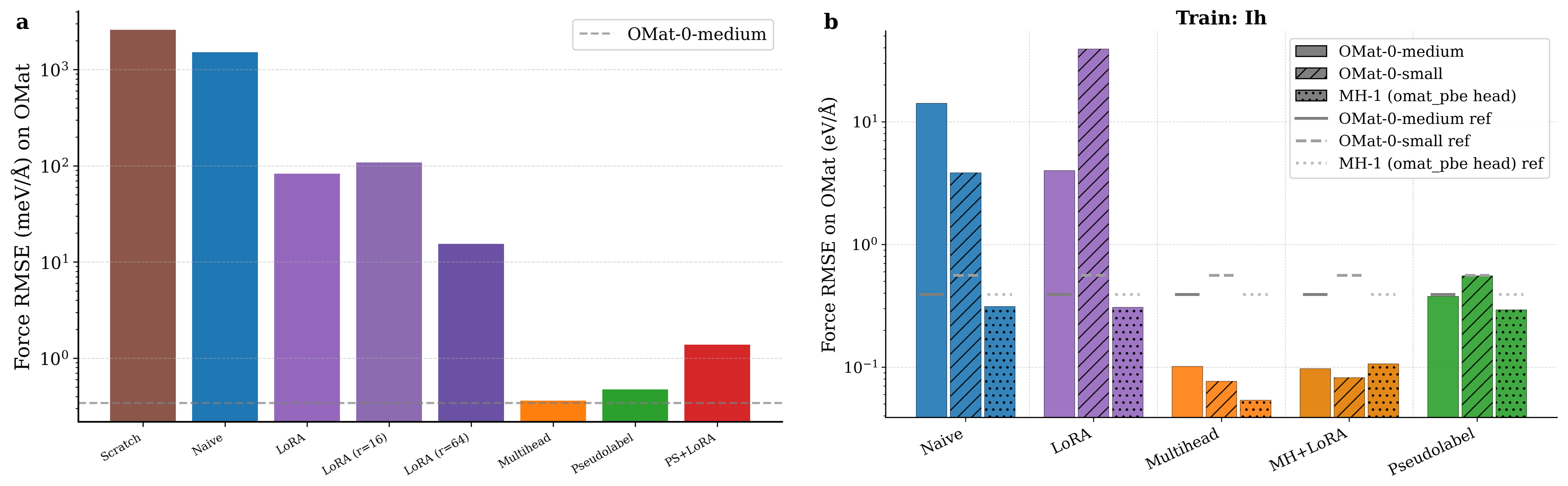}
  \caption{Foundation-model quality preservation after fine-tuning. \textbf{a}, SPICE forgetting with 1 configuration per molecule, measured as force RMSE on the OMat test set after fine-tuning. \textbf{b}, Forgetting for ice polymorph fine-tuning, shown for the Ih-trained models across foundation sizes. Replay-based methods best preserve the pretrained distribution in both settings.}
      \label{fig:foundation_preservation}
\end{figure}

We examine these two requirements in turn. Retaining accuracy on the pretrained distribution is quantified by forgetting. Figure~\ref{fig:foundation_preservation} combines the SPICE forgetting result with the ice-polymorph forgetting analysis for the Ih-trained models. Forgetting is most severe in the SPICE setting: with 1 configuration per molecule, Naive fine-tuning degrades OMat performance by orders of magnitude, while LoRA provides intermediate protection and replay-based methods maintain near-foundation accuracy. In both cases, Naive and LoRA move furthest away from the pretrained distribution, whereas Multihead and related replay methods preserve the foundation model substantially better.

The second requirement, a physically sensible repulsive wall, requires diagnostics beyond standard validation metrics. Standard validation metrics probe interpolation on data drawn from the training distribution, but they cannot detect pathological failures in unexplored regions of configuration space. A fine-tuned model may achieve low validation errors while developing unphysical ``holes'' in its PES: regions where the model predicts strong attraction instead of the steep repulsion that should arise when atoms are pushed into close proximity. To probe for these defects systematically, we employ an RSS protocol designed to sample diverse local minima across the PES and identify failures in the repulsive wall; further details are given in Appendix~\ref{sec:rss}. All models include the standard MACE explicit pair-repulsion term used by the OMat models, so these holes do not arise from the absence of a pairwise short-range prior. This term is a Ziegler--Biersack--Littmark (ZBL) screened Coulomb pair potential, added directly to the total energy and smoothly tapered to zero with a polynomial cutoff at the sum of the covalent radii of the two atoms. The RSS failures therefore indicate breakdowns in the learned many-body contribution to the repulsive wall, rather than a missing explicit pair repulsion.

\begin{figure}[h]
    \centering
    \includegraphics[width=0.9\textwidth]{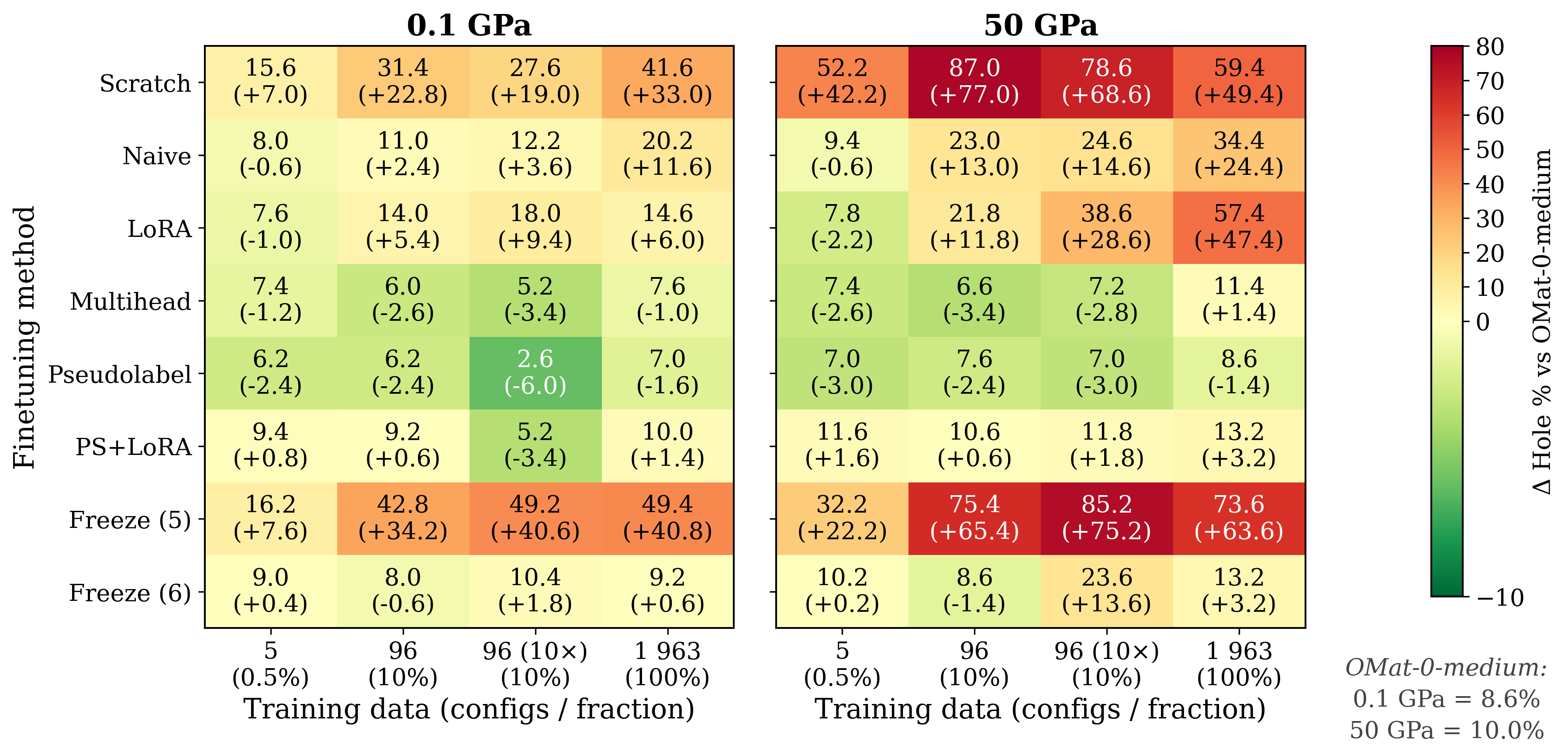}
    \caption{Random structure search summary for fine-tuned NaCl models. Heat maps report the fraction of flagged relaxed structures under the RSS criteria described in Appendix~\ref{sec:rss}.}
    \label{fig:nacl_rss}
\end{figure}

Our results reveal a consistent ordering of methods by robustness. Figure~\ref{fig:nacl_rss} shows that replay-based methods preserve this behaviour more reliably, while naive fine-tuning, LoRA, and from-scratch training introduce a larger number of flagged structures. Multihead replay models and pseudolabel replay models retain the foundation model's robustness, producing few or no flagged structures, with pseudolabel replay showing a slight advantage. Freezing all layers except the readout also preserves repulsive wall integrity, consistent with the fact that the embedding and message-passing representations responsible for short-range interactions remain unchanged. Naive fine-tuning and LoRA introduce a significant number of PES holes. Training from scratch and partial freezing strategies that leave early interaction layers trainable produce the most holes, indicating that unconstrained modification of the interaction layers is particularly damaging to the repulsive wall.

A notable secondary finding is that the number of PES holes increases with training set size for methods trained without replay: models trained on more data develop more holes in the repulsive wall, even as their in-domain accuracy improves. This is consistent with the interpretation that longer or more intensive training on a narrow distribution drives the model further from its pretrained solution. Multihead replay methods do not exhibit this trend, confirming that the replay objective preserves repulsive wall integrity even as target-task training intensifies.

These results provide evidence that fine-tuning without replay degrades the model in regions of configuration space not represented in the fine-tuning data. A complementary comparison of replay-set choices (element-matched OMat24, random MPTraj, and the combination) for the same RSS metric is reported in Appendix~\ref{app:nacl_replay} (Fig.~\ref{fig:nacl_replay_si}). At the same time, the downstream consequences of these holes are context-dependent. For the systems and conditions tested here, including MD simulations at moderate temperatures, the PES holes do not cause observable instabilities: trajectories remain stable and thermodynamic properties are well reproduced despite the presence of holes detectable by RSS.
The need for replay is therefore tied to deployment scope: narrow applications can tolerate local adaptation, while broad searches require preservation of the foundation model's repulsive wall.

\section{Discussion}
Fine-tuning foundation models can achieve as good or better accuracy on target tasks as models trained from scratch, using less data, with the added benefit of inheriting some of the broad robustness of a model trained on a diverse set of materials. Here we have looked at what determines how well fine-tuning performs, and how much of the  foundation model's robustness is preserved depending on the fine-tuning method.
The results support a simple hierarchy: foundation model quality, E0 consistency, and stable optimisation must be controlled before method-level differences can be interpreted. When any of these prerequisites is misconfigured, it can dominate the outcome and obscure the behaviour of the fine-tuning method itself. This suggests that some of the difficulty reported in early MLIP fine-tuning studies may have reflected weaker foundations, inconsistent reference energies, or unstable training protocols rather than an intrinsic failure mode of naive fine-tuning. 
Once these prerequisites are controlled, method choice is best understood in terms of deployment scope. For a narrow target application, the fine-tuned model only needs to be accurate on configurations representative of the target system. In this regime, naive fine-tuning is difficult to beat: it uses the full model capacity, converges efficiently, and achieves target-task accuracy comparable to or better than the constrained alternatives.

For broader deployment, the relevant objective changes. A model used for screening, structure prediction, or transfer across a materials family must preserve the foundation model's stable behaviour away from the fine-tuning distribution. In our benchmarks, replay-based methods are the only approaches that consistently maintain near-foundation accuracy on the pretraining distribution and preserve the short-range repulsive wall under RSS. The RSS holes produced by naive and LoRA fine-tuning did not cause observable instabilities in the moderate-temperature MD simulations tested here, but they are still important for applications that deliberately explore unusual or high-energy configurations.

Within the replay family, replay labels and replay structures appear to have separable effects. When the replay structures are held fixed, original-label and pseudolabel replay give nearly equivalent results, showing that access to the original DFT replay labels is not essential for maintaining foundation-model behaviour. The replay-source comparison in Appendix~\ref{app:replay_data_ablation} shows the complementary result: for narrow target-task accuracy, element-matched OMat24 replay, random MPTraj replay, and their combination give essentially indistinguishable errors when all are pseudolabelled. In the lithium electrolyte benchmark, the broad replay set of 10,000 randomly sampled MPTraj structures with OMat24 pseudolabels improves out-of-distribution performance more than fine-tuning without replay, despite the replay structures coming from outside the OMat24 pretraining corpus. These results suggest that broad structural coverage is more important than replay-label source for model breadth maintenance. 

LoRA occupies an intermediate role. For MLIPs its value is not primarily parameter efficiency---MACE foundation models are small relative to the cost of storing atomic graphs---but capacity control: low ranks restrict departure from the pretrained solution, while higher ranks recover the flexibility needed for diverse tasks such as SPICE. Empirically this effect is present but modest: higher ranks improve target-task accuracy on broader datasets (most clearly SPICE, see Section~\ref{sec:method_comparison}), and LoRA reduces forgetting slightly relative to naive fine-tuning, but neither effect approaches the magnitude of multihead replay. This also clarifies the contrast between LoRA and layer freezing. Layer freezing is a depth-wise restriction of adaptability: early layers are held fixed while later layers retain full flexibility. LoRA is a width-wise restriction: all targeted layers can change, but only within a reduced-rank subspace. Our results suggest that the width-wise constraint is generally more effective for MACE, consistent with the idea that useful chemical representations are distributed across interaction layers rather than confined to the last layers.

Fine-tuning shows clear data-efficiency advantages over training from scratch in the low-data comparisons considered here, although the form of the advantage differs by benchmark. For aqueous NaCl, fine-tuned models recover broadly correct solvation structure already with 10\% of the training data, whereas the from-scratch model only approaches a reasonable RDF at the full dataset size. For S$_\mathrm{N}$2, the from-scratch model trained on the same 60 reactant/product configurations substantially underestimates the reaction barrier; in the original study, accurate barriers required active-learning iterations that added transition-state-region configurations to the training set~\cite{kuryla2025efficient}. For SPICE, naive fine-tuning outperforms the matched from-scratch baseline across all tested training-set fractions and subsets. Together these results indicate that pretrained representations reduce the amount and diversity of new reference data needed to obtain useful target-task models.

The cost of preserving breadth is additional training complexity. Naive fine-tuning, layer freezing, and LoRA have similar per-epoch costs, whereas multihead replay processes both target and replay data and requires a lower learning rate for stable convergence (Appendix Fig.~\ref{fig:mh_lr_ablation}). The combination of more data per epoch and slower convergence per epoch means that multihead training requires approximately 3-fold to 15-fold more training compute than naive fine-tuning, depending on the dataset sizes. This computational overhead is a practical consideration that must be weighed against the breadth-maintenance benefits.

The scope of this study leaves some questions open. First, all experiments use the MACE architecture; while the physical inductive biases (equivariance, message passing) are shared across many MLIP architectures, the specific behaviour of LoRA, layer freezing, and multihead training may differ for different architectures.
Generalisation of our findings to other architectures remains to be tested.

Second, the hyperparameter optimisation was necessarily limited. We tuned the principal settings for each method, including learning rate, energy and force weights in the loss, batch size, EMA decay, gradient clipping, and weight decay, but did not attempt an exhaustive method-specific search. More extensive optimisation could change some relative rankings, particularly for LoRA, where performance may depend on which layers receive adapters, and for multihead replay, where the replay-to-target data ratio and loss weighting introduce additional degrees of freedom.

Third, our RSS evaluation of PES holes, while systematic, does not exhaustively cover all possible failure modes. The relax–rattle protocol probes short-range repulsive wall failures, but other types of PES artefacts (e.g. incorrect relative stabilities of competing phases) may require different diagnostic approaches.

Finally, the foundation models tested here were trained predominantly on periodic inorganic systems. The SPICE results show that effective fine-tuning is possible even from such foundations, but the present benchmarks do not exhaustively cover the breadth of molecular applications.

\section{Conclusion}

We implemented three capabilities in the MACE codebase---LoRA,
pseudolabelled replay, and model-aware E0 reestimation---and used them, alongside existing
approaches, to benchmark seven fine-tuning strategies across our chemically diverse
benchmark suite. Successful fine-tuning depends first on foundation-model quality, consistent atomic reference energies, and stable optimisation; once these prerequisites are controlled, naive fine-tuning is a strong default for narrow target applications, while replay-based methods are preferable when broad out-of-distribution robustness is required.

Key practical recommendations:
\begin{enumerate}
    \item \textbf{Start from the strongest available foundation model.} In these benchmarks, OMat24-based foundations outperform MPTraj-based foundations by more than the differences between most fine-tuning methods.
    \item \textbf{Use consistent atomic reference energies.} Compute isolated-atom energies at the target level of theory where possible, or use the model-based reestimation procedure described in Section~\ref{sec:e0_method}. Avoid training-set averaged E0s.
    \item \textbf{Optimise each method with stable hyperparameters.} Use zero weight decay for fine-tuning, method-appropriate learning rates, and constant target-task loss weights rather than multistage loss schedules.
    \item \textbf{Use naive fine-tuning as the first strong baseline for narrow applications.} When deployment is restricted to configurations representative of a single target system, naive fine-tuning is accurate, efficient, and requires no additional infrastructure. If naive fine-tuning produces instabilities or unphysical behaviour in MDs, multihead replay should be the default fallback.
    \item \textbf{Use multihead replay when robustness is key.}  Multihead replay, with either original or pseudolabelled replay data, preserves foundation-model breadth and repulsive-wall integrity when the model may encounter configurations outside the fine-tuning distribution. 
\end{enumerate}
The near-equivalence of pseudolabelled replay and original-label replay in target-task accuracy effectively decouples the replay structures from the original pretraining corpus: any structurally diverse dataset can serve as the replay source. This raises a question that the present work does not address: what is the optimal composition of the replay set? In particular, which structures, in what relative proportions, and at what total set size most efficiently preserve foundation-model breadth, and how should this composition depend on the target task and on the foundation model's pretraining distribution? We leave a systematic investigation of optimal replay composition to future work.

A further direction is to test how more substantial architectural advances in foundation MLIPs affect fine-tuning behaviour. The foundation-model comparison in this work mainly probes improvements in pretraining data scale, data diversity, and model capacity. Recent developments such as MACE-POLAR-1~\cite{mace_polar_1}, which augments MACE with explicit long-range electrostatics, polarisable charge and spin densities, and global charge and spin constraints, represent a more significant change to the model architecture and provide a natural next test case. More broadly, future foundation MLIPs may differ not only through larger datasets or more accurate energy, force, and stress labels, but also through richer input information and auxiliary physical descriptors, such as total charge, total spin, partial charges, or electronic descriptors including density-of-states information.

\section{Data Availability}

Training datasets, model checkpoints, and evaluation scripts are available at \url{https://huggingface.co/datasets/ev-tlt/MACE_finetuning_supplementary}.

\section{Code Availability}
The MACE codebase, including the LoRA fine-tuning, pseudolabel replay, and E0 reestimation tools described in this work, is available at \url{https://github.com/ACEsuit/mace} (MACE version 3.15+). 

\section{Acknowledgements}

E.V.U. acknowledges support from the EPSRC Centre for
Doctoral Training in Automated Chemical Synthesis Enabled by
Digital Molecular Technologies (SynTech) with Grant Reference
No. EP/S024220/1. A.M.E. was supported by the Ada Lovelace Centre at the Science and Technology Facilities Council \\(https://adalovelacecentre.ac.uk/), the Physical Sciences Data Infrastructure (https://psdi.ac.uk; jointly STFC and the University of Southampton) under grants EP/X032663/1 and EP/X032701/1, and EPSRC under grants EP/W026775/1 and EP/V028537/1. The work of N.B. was supported by the U.~S. Naval Research Laboratory's fundamental research base program.
The bulk of the computations in this work were performed using the Isambard-AI National AI Research Resource (AIRR), operated by the University of Bristol and funded by the UK Government's Department for Science, Innovation and Technology (DSIT) via UK Research and Innovation and the Science and Technology Facilities Council [ST/AIRR/I-A-I/1023]~\cite{isambard_ai}. We also acknowledge computational resources and support from the Cambridge Service for Data Driven Discovery (CSD3), the Max Planck Computing and Data Facility (MPCDF), including the Viper-GPU system based on AMD Instinct MI300A APUs, and STFC Scientific Computing Department's SCARF cluster. I.B. was supported by the Harding
Distinguished Postgraduate Scholarship.

\section{Author Contributions}

T.L.T.\ and E.V.U.\ contributed equally to this work. T.L.T., E.V.U., I.B.\ and G.C.\ conceived the study and designed the methodology. T.L.T.\ and E.V.U.\ developed the software, performed the investigations, curated the data and produced the figures. A.M.E.\ and N.B.\ contributed to the investigations and data curation. T.L.T., E.V.U., I.B., A.M.E., N.B.\ and G.C.\ carried out the formal analysis. G.C.\ and N.B.\ supervised the work. All authors contributed to reviewing and editing the manuscript.

\section{Competing Interests}

GC is a partner in Symmetric Group LLP that licenses force fields commercially and also has equity interest in Ångström AI.

\bibliographystyle{naturemag}
\bibliography{sample}

\clearpage
\appendix
\renewcommand{\thesection}{\Alph{section}}
\section*{Appendix}

This appendix collects the computational details that support the main text and the additional results and comparisons referenced from it. The first group of subsections (\textit{Foundation models and computational setup} through \textit{Computational resources}) documents the models, loss, training procedure, and benchmark systems. The remaining subsections report supplementary results and comparisons: foundation-model capacity in ice cross-phase learning, single-phase versus combined ice fine-tuning, the full NaCl RDF panel, the NaCl replay-strategy comparison, MACE-MH1 isolated-atom RMSE (reduced and stress-included variants), the loss-weight schedule comparison, the multihead replay learning-rate comparison, and the replay-data composition comparison.

\section{Foundation models and computational setup}
\label{app:foundation_setup}

\begin{table}[h]
\centering
\begin{tabular}{lccccr}
\toprule
\textbf{Model} & \textbf{Hidden irreps} & $\boldsymbol{L_{\max}}$ & \textbf{Interaction block} & \textbf{Pair rep.} & \textbf{Params.} \\
\midrule
MACE-OMat-small & $128 \times 0e$ & 0 & DensityResidual & Yes & 8.2M \\
MACE-OMat-0-medium & $128 \times 0e + 128 \times 1o$ & 1 & DensityResidual & Yes & 9.1M \\
MACE-MH1 & $512 \times 0e + 512 \times 1o$ & 1 & ResidualNonLinear & Yes & 6.4M \\
\midrule
MACE-MP0a-medium & $128 \times 0e + 128 \times 1o$ & 1 & AgnosticResidual & No & 4.7M \\
\bottomrule
\end{tabular}
\caption{Architecture summary for foundation models used in this work. All OMat models use Agnesi distance transforms; MACE-MP0a-medium uses no distance transform. MACE-MH1 uses 6 readout heads (one per training dataset: OMat24, OMol25, MatPES, OC20, MP-PBE, SPICE) while other models use a single head. The lower parameter count of MACE-MH1 relative to the OMat models despite its larger channel dimension reflects the use of a different, more compact interaction block architecture.}
\label{tab:models}
\end{table}

The architecture parameters of the foundation models used throughout this work are summarised in Table~\ref{tab:models}. Unless otherwise specified, MACE-OMat-0-medium serves as the primary foundation model for our evaluations, providing a balance between expressivity and computational efficiency. To assess the impact of foundation quality on fine-tuning outcomes, we also compare against MACE-MP0a-medium~\cite{mace_mp0}, pretrained on the earlier MPTraj dataset, which is less chemically diverse and less consistent than OMat24.

For baseline comparisons, we train models from scratch with architectures matching the expressivity-related hyperparameters of the corresponding foundation models (number of channels, $L_{\max}$, hidden irreps, interaction layers). All OMat foundation models were originally trained with Agnesi transforms enabled; in some from-scratch runs, however, Agnesi transforms produced severe optimisation instabilities, so we disabled them for scratch baselines to ensure a fair comparison.

\subsection{Loss function and training objective}
\label{app:loss_function}

The MACE model predicts total energies $E$, atomic forces $\mathbf{F} = -\nabla_{\mathbf{R}} E$, and optionally stress tensors $\boldsymbol{\sigma}$ for a given atomic configuration. We train using a combined loss function:
\begin{equation}
\mathcal{L} = \lambda_E \mathcal{L}_E + \lambda_F \mathcal{L}_F + \lambda_S \mathcal{L}_S
\end{equation}
where the stress term is included only for datasets with stress labels (e.g., ice polymorphs; $\lambda_S = 1.0$). The individual losses use the Huber function, which behaves as $L^2$ for small residuals and transitions to $L^1$ for large residuals, reducing sensitivity to outliers:
\begin{equation}
\operatorname{Huber}_\delta(x) = \begin{cases} x^2/2 & \text{if } |x| \leq \delta \\ \delta(|x| - \delta/2) & \text{otherwise} \end{cases}
\end{equation}
The energy, force, and stress losses are then:
\begin{align}
\mathcal{L}_E &= \frac{1}{N_{\text{batch}}} \sum_{i=1}^{N_{\text{batch}}} \operatorname{Huber}_\delta\bigl(E_i^{\text{pred}} - E_i^{\text{ref}}\bigr) \\
\mathcal{L}_F &= \frac{1}{N_{\text{batch}}} \sum_{i=1}^{N_{\text{batch}}} \frac{1}{N_{\text{atoms},i}} \sum_{j=1}^{N_{\text{atoms},i}} \operatorname{Huber}_\delta\bigl(\,\| \mathbf{F}_{ij}^{\text{pred}} - \mathbf{F}_{ij}^{\text{ref}} \|\,\bigr) \\
\mathcal{L}_S &= \frac{1}{N_{\text{batch}}} \sum_{i=1}^{N_{\text{batch}}} \operatorname{Huber}_\delta\bigl(\,\| \boldsymbol{\sigma}_{i}^{\text{pred}} - \boldsymbol{\sigma}_{i}^{\text{ref}} \|_\mathrm{F}\,\bigr)
\end{align}
We use the MACE default Huber transition parameter $\delta = 0.01$ for all loss terms, matching the OMat24 pretraining configuration. Stress errors follow trends consistent with force errors across all methods and are not discussed separately. For the target task, we use $\lambda_E = 10, \lambda_F = 10$ throughout training; for multihead replay, the replay head uses $\lambda_E = 1, \lambda_F = 10$, matching the pretraining objective. When stress labels are available, we use $\lambda_S = 1.0$.

\subsection{Combined approaches}
\label{app:combined}

We also evaluate \textbf{multihead\,+\,LoRA}, combining replay training with low-rank constraints by applying LoRA adapters to the shared representation network while maintaining separate readout heads. This combination appears in only one scenario in our evaluation: at extreme data scarcity (5 training configurations for S$_\mathrm{N}$2 reactions), where standard multihead training fails but multihead\,+\,LoRA succeeds. We hypothesise that at extreme scarcity, the replay and target objectives provide conflicting signals, but the LoRA constraint filters the model's response by restricting how it can respond to competing gradients. In all other scenarios tested, the combined regularisation of replay\,+\,LoRA produces worse results than either approach individually, consistent with over-regularisation.

\subsection{Benchmark dataset summary}
\label{app:benchmarks}

\begin{table}[h]
\centering
\begin{tabular}{p{0.14\textwidth} p{0.15\textwidth} p{0.20\textwidth} p{0.24\textwidth} p{0.17\textwidth}}
\toprule
\textbf{System} & \textbf{Reference level} & \textbf{Fine-tuning data} & \textbf{Held-out / downstream evaluation} & \textbf{Replay data} \\
\midrule
Aqueous NaCl & MP2~\cite{oneill2024pairpairmachinelearnedexplicitlycorrelated} & 963 configurations, with data fractions used for scaling experiments & RDFs from MD trajectories, compared against the BPNN reference & 2\,912 element-matched foundation-data replay configurations \\
S$_\mathrm{N}$2 reactions & MP2~\cite{kuryla2025efficient} & 60 reactant/product configurations, excluding the transition state & NEB barrier profiles and pointwise energy/force errors & 3\,706 element-matched foundation-data replay configurations \\
Ice polymorphs & PBE+D3 & 400 configurations across phases Ih, II, VI, and VIII (100 per phase)~\cite{kaur2024dataefficientfinetuningfoundationalmodels} & Separate held-out test configurations for the water/ice phases; cross-phase force errors and MD stability tests & 196 element-matched foundation-data replay configurations \\
SPICE biomolecules & $\omega$B97M-D3(BJ)/def2-TZVPPD~\cite{spice,MACEOFF} & 951\,005 training/validation configurations from the cleaned MACE-OFF subset & Separate 50\,000-configuration test set; subset-resolved energy/force errors and TorsionNet500 torsion profiles & 39\,459 element-matched foundation-data replay configurations \\
Lithium electrolytes & PBE & 500 LPSC configurations (Li$_6$PS$_5$Cl, single argyrodite composition) for fine-tuning and validation & Broad held-out electrolyte evaluation: over 130 other argyrodite compositions and several non-argyrodite structures, all unseen during fine-tuning & Pseudolabelled replay using a 10\,000-configuration random MPTraj subsample \\
\bottomrule
\end{tabular}
\caption{\textbf{Benchmark systems used for fine-tuning and evaluation.} The table separates the data used for fine-tuning and validation from held-out or downstream evaluations, and summarises the replay data used by replay-based methods. SPICE and the water/ice benchmark use separate test sets, while the lithium electrolyte benchmark includes a broad out-of-distribution test set spanning many compositions and structures beyond LPSC.}
\label{tab:benchmark_systems}
\end{table}

\subsection{Hyperparameter summary}
\label{app:hypers}

Table~\ref{tab:hyperparams} summarises the complete training configuration for all methods evaluated. Each method's hyperparameters were individually optimised to achieve stable training dynamics and well-converged models.
All models were trained using the AdamW optimiser~\cite{adamW} with default momentum parameters ($\beta_1 = 0.9$, $\beta_2 = 0.999$). Most training was performed on NVIDIA GH200 GPUs, with the lithium models trained on NVIDIA A100 GPUs and additional runs carried out on AMD Instinct MI300A APUs on the MPCDF Viper-GPU system. For datasets that include stress tensors (e.g., the ice polymorph system), stress matching is included in the loss function; stress errors follow trends consistent with force errors across all methods and are not discussed separately.

\begin{table}[h]
\centering
\begin{tabular}{lcccc}
\toprule
\textbf{Method} & \textbf{Learning Rate} & \textbf{EMA Decay} & \textbf{Grad Clip} & \textbf{Trainable} \\
\midrule
From-scratch & $10^{-2}$ & 0.99 & 10.0 & 100\% \\
Naive & $10^{-3}$ & 0.999 & 1.0 & 100\% \\
Freeze & $10^{-3}$ & 0.999 & 1.0 & $\sim$5\% \\
LoRA ($r=4$) & $10^{-2}$ & 0.99 & 10.0 & $\sim$2.5\% \\
LoRA ($r=16$) & $10^{-2}$ & 0.99 & 10.0 & $\sim$10\% \\
LoRA ($r=64$) & $10^{-2}$ & 0.99 & 10.0 & $\sim$30\% \\
Multihead & $10^{-4}$ & 0.9999 & 1.0 & 100\% \\
Pseudolabel MH & $10^{-4}$ & 0.9999 & 1.0 & 100\% \\
MH + LoRA & $10^{-3}$ & 0.999 & 10.0 & $\sim$2.5\% \\
\bottomrule
\end{tabular}
\caption{Hyperparameter configurations for all fine-tuning methods. Each row reflects per-method tuning to converged, stable training.}
\label{tab:hyperparams}
\end{table}

\subsection{Random structure search protocol}
\label{sec:rss}

We generate random crystal structures using the PyXtal package~\cite{pyxtal} for 10 fixed stoichiometries, each comprising two or three distinct atomic species drawn from the elements present in the fine-tuning dataset. For each composition, 50 independent initial configurations are created, yielding 500 structures per model. Each structure is then subjected to an iterative relax--rattle loop: the structure is relaxed with the model, random perturbations are applied to the atomic positions, and the perturbed structure is relaxed again. This cycle is repeated three times, encouraging the optimiser to explore multiple basins and increasing the likelihood of encountering pathological regions of the PES.

The relaxed structures are then screened for signatures of PES holes using the following criteria:
\begin{enumerate}
    \item \textbf{Atomic overlap:} structures where the minimum interatomic-distance ratio is below 0.5, i.e., the closest interatomic distance is less than 50\% of the sum of the corresponding covalent radii.
    \item \textbf{Anomalous volume:} structures with volume per atom below 30\% of the median volume per atom for that composition.
    \item \textbf{Anomalous energy:} structures with energy per atom more than 10 median absolute deviations (MADs) below the median energy per atom for that composition, with a minimum threshold of 5~eV/atom.
    \item \textbf{Combined volume and energy with relaxed thresholds:} structures with volume per atom below 50\% of the composition median and energy per atom below the composition median.
\end{enumerate}

Each of these signatures reflects the same underlying failure mode: the model's PES contains regions where the physically mandated short-range repulsion has been replaced by strong attraction, causing structures to implode to unphysically dense, low-energy states during relaxation. We apply this protocol to both NaCl-fine-tuned and lithium-electrolyte-fine-tuned models to test whether successful fine-tuning on narrow target distributions preserves physically sensible short-range behaviour.
\subsection{Computational resources}
\label{app:resources}

Training hardware assignments are summarised in Appendix~\ref{app:foundation_setup}, and facility acknowledgements are given above. Full training configurations are detailed in Table~\ref{tab:hyperparams}.
\clearpage

\section{MACE-MH1 isolated-atom RMSE with MACE stress loss}
\label{app:mh1_stress}

For completeness, the MACE-MH1 isolated-atom RMSE results reported in Table~\ref{tab:universal_data} (Universal/Huber loss configuration) are supplemented here by the corresponding values using MACE's stress loss, i.e. a weighted mean-squared-error objective over energy, forces, and stress (Table~\ref{tab:stress_data}). The columns follow the same convention as Table~\ref{tab:universal_data}. Isolated-atom configurations had a weight of 1000, while all other configurations had a weight of 1. The qualitative trends are the same as for the Universal/Huber configuration.

\begin{table}[hbt!]
\centering
\begin{tabular}{l c c c c c c}
\hline
\multirow{2}{*}{Repeats} & \multicolumn{4}{c}{Multihead (DFT replay labels)} & \multicolumn{2}{c}{Pseudolabel} \\
\cmidrule(lr){2-5} \cmidrule(lr){6-7}
            & DFT E0s & Reest.\ E0s & DFT E0s (2-stage) & Reest.\ E0s (2-stage) & DFT E0s & Reest.\ E0s \\
\hline
1           & 9.0    & 18.2   & 3.2   & 4.7   & 20.3   & 24.7   \\
5           & 2.3    & 7.3    & 2.0   & 1.2   & 3.5    & 3.1    \\
10          & 1.9    & 4.3    & 0.5   & 1.0   & 3.2    & 4.4    \\
20          & 1.4    & 3.5    & 0.5   & 0.6   & 1.9    & 2.8    \\
40          & 1.2    & 1.3    & 0.4   & 0.2   & 1.3    & 1.1    \\
\hline
\end{tabular}
\caption{\textbf{Isolated-atom RMSE for MACE-MH1 multihead fine-tuning under MACE's stress loss.} Same configurations as Table~\ref{tab:universal_data}, using a weighted mean-squared-error objective over energy, forces, and stress. All values are in meV/atom; lower is better. Columns marked ``2-stage'' use a two-stage loss schedule that raises the energy weight relative to the force weight partway through training; all other columns use constant loss weights throughout.}
\label{tab:stress_data}
\end{table}

\section{Loss-weight schedule comparison}
\label{app:loss_schedule_ablation}

Two-stage loss schedules---first emphasising force matching, then transitioning to energy matching---are common in from-scratch MLIP training~\cite{medicineMACE}. To explain this empirical observation it has been speculated that the initially higher force weights may be more effective at steering the fit to a general region in parameter space that contains good solutions to the fit, but once the parameters are in that region, the higher energy weights are more effective at steering the fit to regions with improved energy accuracy without appreciably degrading force accuracy. However, this rationale does not apply to fine-tuning, where the parameter values of the pretrained model are already in a region that is likely to have low loss minima as a result of the pretraining itself. In fact, it appears that changing the weights during fine-tuning interferes with the progress of the loss minimisation. 
 
\begin{figure}[h]
    \centering
    \includegraphics[width=0.95\textwidth]{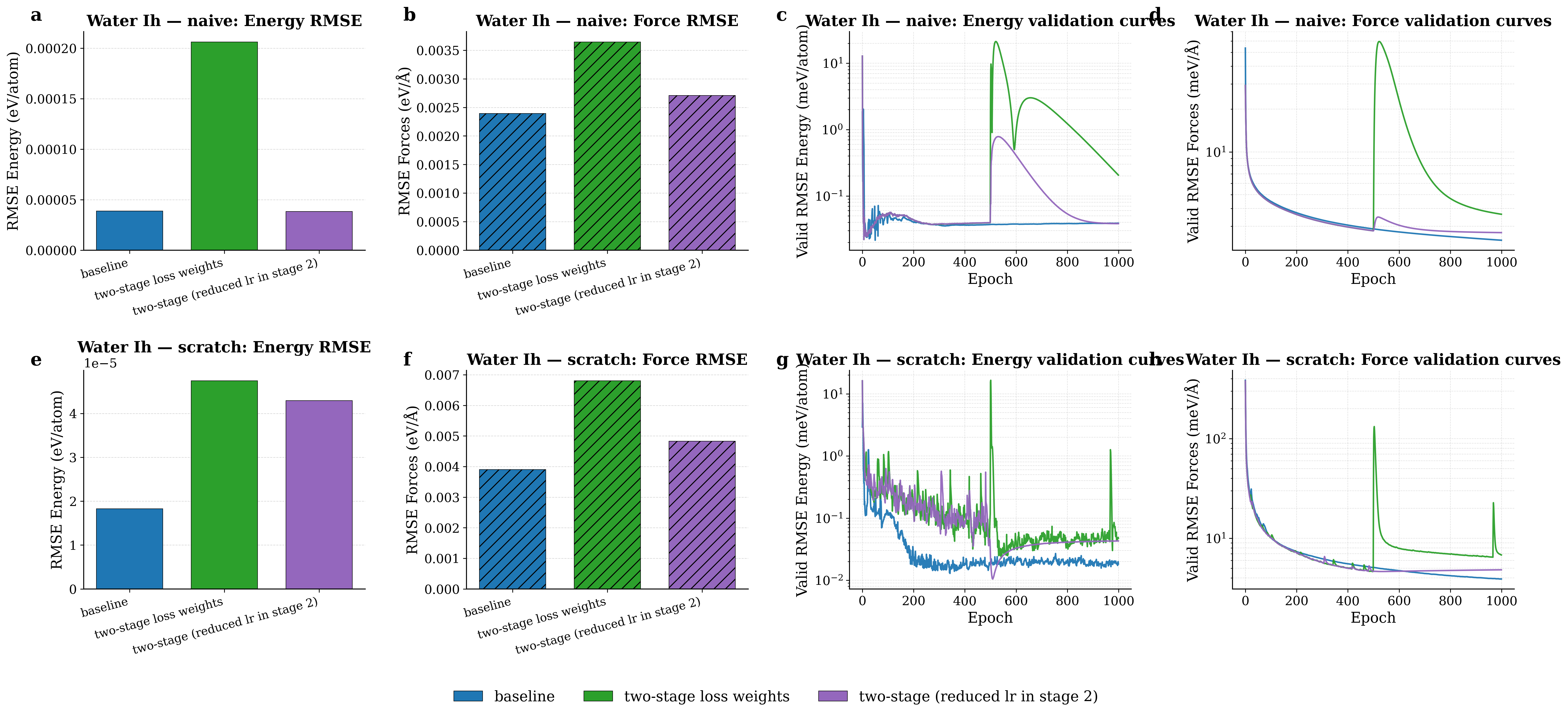}
   \caption{\textbf{Loss-weight schedule comparison on ice Ih.} Comparison of constant target-task loss weights ($\lambda_E = 10$, $\lambda_F = 10$ throughout training, ``baseline'') against two-stage schedules in which $\lambda_F$ is dominant in stage one and $\lambda_E$ is increased in stage two (starting at epoch 500), with and without a reduced learning rate in stage two. \textbf{a,b,c,d}, Naive fine-tuning of MACE-OMat-0-medium: final validation energy and force RMSE relative to the baseline (a) and validation error curves over training (b). \textbf{e,f,g,h}, From-scratch training on the same ice Ih dataset: corresponding bars and error curves. The staged schedules introduce a sharp transient at the stage-two transition and do not improve over the constant-weight baseline in either setting.}
     \label{fig:loss_schedule_ablation}
\end{figure}

Figure~\ref{fig:loss_schedule_ablation} compares the constant-weight configuration used throughout this work ($\lambda_E = 10$, $\lambda_F = 10$ for the entire run) against representative two-stage schedules under otherwise identical training settings, for both naive fine-tuning of MACE-OMat-0-medium and from-scratch training on the ice Ih system. In the naive fine-tuning case (Fig.~\ref{fig:loss_schedule_ablation}a, b), the two-stage schedule with the standard learning rate causes a sharp degradation of both energy and force errors at the stage-two transition (epoch 500), from which the model only partially recovers; reducing the stage-two learning rate suppresses the spike but still does not improve over the constant-weight baseline. From-scratch training (Fig.~\ref{fig:loss_schedule_ablation}c, d) shows the same qualitative behaviour: the staged schedules destabilise training at the transition and yield higher final errors than the constant-weight configuration. We therefore use constant target-task loss weights throughout the experiments reported in the main text.

\clearpage
\section{Multihead replay learning-rate comparison}
\label{app:mh_lr_ablation}

\begin{figure}[h]
    \centering
    \includegraphics[width=0.95\textwidth]{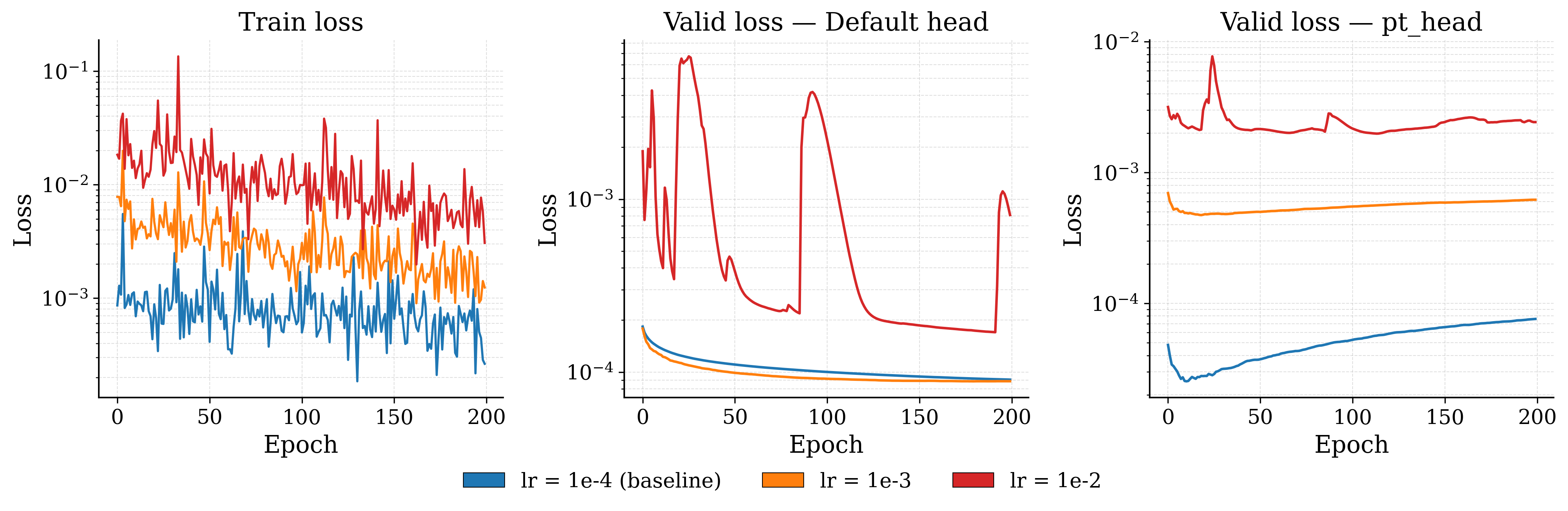}
   \caption{\textbf{Multihead replay is sensitive to the learning rate (LPSC).} Multihead replay fine-tuning of MACE-OMat-0-medium on the lithium electrolyte (LPSC) system with pseudolabelled MPTraj replay, swept over learning rates $\{10^{-4},\,10^{-3},\,10^{-2}\}$. Training and validation loss curves for the target head (``Default'') and the replay head (``pt\_head'') as a function of epoch. At the higher learning rates the loss on the replay data immediately jumps to a high value and never recovers. Final-epoch validation metrics for both heads are reported in Table~\ref{tab:mh_lr_rmse}.
    }
     \label{fig:mh_lr_ablation}
\end{figure}

\begin{figure}[h]
    \centering
    \includegraphics[width=0.95\textwidth]{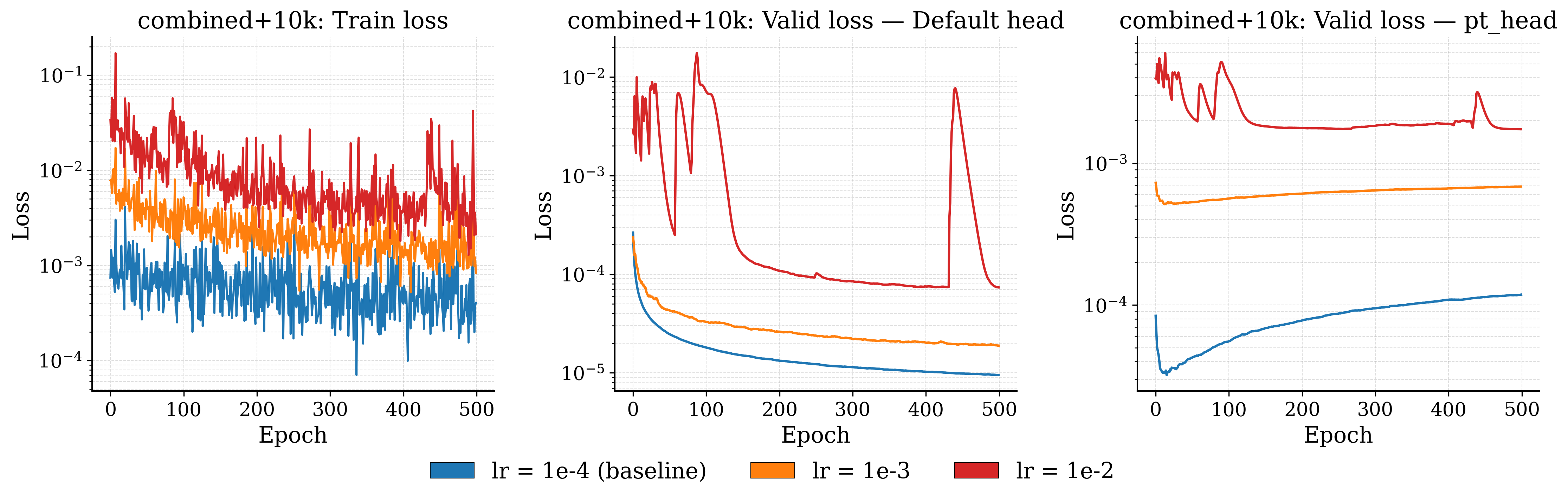}
   \caption{\textbf{Multihead replay learning-rate sensitivity on the S$_\mathrm{N}$2 system.} Multihead replay fine-tuning of MACE-OMat-0-medium on the S$_\mathrm{N}$2 reaction dataset, swept over learning rates $\{10^{-4},\,10^{-3},\,10^{-2}\}$ for two pseudolabelled replay-set choices: an OMat24 element-matched subsample and a combined OMat24 + 10\,000-structure MPTraj subsample. Training and validation losses for the target head (``Default'') and replay head (``pt\_head'') as a function of epoch, shown for each head choice. Final-epoch validation metrics are reported in Table~\ref{tab:mh_lr_rmse}.}
     \label{fig:mh_lr_ablation_sn2}
\end{figure}

The hyperparameter table (Table~\ref{tab:hyperparams}) reports a learning rate of $10^{-4}$ for multihead replay training, an order of magnitude lower than for naive fine-tuning ($10^{-3}$) and two orders lower than for LoRA ($10^{-2}$). To document the dependence behind this choice, we sweep the learning rate over $\{10^{-2}, 10^{-3}, 10^{-4}\}$ for multihead replay fine-tuning of MACE-OMat-0-medium on the lithium electrolyte (LPSC) system, holding all other settings fixed and using a 10\,000-structure random MPTraj subsample as the pseudolabelled replay set. Figure~\ref{fig:mh_lr_ablation} shows the training dynamics for the target (``Default'') head and the replay (``pt\_head'') head, and Table~\ref{tab:mh_lr_rmse} reports the corresponding final-epoch validation metrics. The training and validation losses (Fig.~\ref{fig:mh_lr_ablation}) decrease smoothly and to substantially lower values for $10^{-4}$ than for $10^{-3}$ and $10^{-2}$, on both heads. The final-epoch summary (Table~\ref{tab:mh_lr_rmse}) makes the same point quantitatively: at $10^{-3}$ the replay-head force RMSE is roughly an order of magnitude higher than at $10^{-4}$ (218 vs 24 meV/\AA), and at $10^{-2}$ both heads diverge from the baseline by orders of magnitude on energy, force and loss. The target-head force RMSE is essentially identical between $10^{-4}$ and $10^{-3}$ but the replay-head accuracy is much worse at $10^{-3}$, indicating that the breadth-maintenance benefit of replay is degraded well before the target task itself shows signs of trouble. The same trend is reproduced on a chemically distinct system in Fig.~\ref{fig:mh_lr_ablation_sn2}, where multihead replay fine-tuning of MACE-OMat-0-medium on the S$_\mathrm{N}$2 reaction dataset is swept over the same three learning rates for two replay-set choices (an OMat24 element-matched subsample and a combined OMat24+MPTraj subsample): $10^{-4}$ again gives the lowest losses on both heads, while $10^{-3}$ already inflates the replay-head force RMSE by roughly an order of magnitude (e.g. 202--224 vs 88--53~meV/\AA) and $10^{-2}$ destabilises both heads. We therefore use $10^{-4}$ as the default for all multihead replay experiments.
\begin{table}[h]
\centering
\caption{Final-epoch validation RMSE for energy (meV/atom) and forces (meV/\AA) from the multihead + pseudolabel learning-rate sweep, reported for the Default head (task) and pt\_head (replay).}
\label{tab:mh_lr_rmse}
\begin{tabular}{llcccc}
\toprule
 &  & \multicolumn{2}{c}{Default head} & \multicolumn{2}{c}{pt\_head (replay)} \\
\cmidrule(lr){3-4} \cmidrule(lr){5-6}
System & Learning rate & RMSE E & RMSE F & RMSE E & RMSE F \\
 &  & (meV/atom) & (meV/\AA) & (meV/atom) & (meV/\AA) \\
\midrule
Li (LPSC) & 1e-4 (baseline) & \textbf{0.5} & 18.4 & \textbf{6.1} & \textbf{23.9} \\
 & 1e-3 & 0.7 & \textbf{18.1} & 41.2 & 217.6 \\
 & 1e-2 & 1.7 & 30.4 & 772.9 & 963.8 \\
\midrule
S$_\mathrm{N}$2 (combined+10k) & 1e-4 (baseline) & \textbf{0.0} & \textbf{4.5} & \textbf{15.2} & \textbf{53.3} \\
 & 1e-3 & 0.5 & 6.7 & 76.6 & 224.2 \\
 & 1e-2 & 0.4 & 16.0 & 230.1 & 459.8 \\
\bottomrule
\end{tabular}
\end{table}

\clearpage
\section{Replay-data composition comparison}
\label{app:replay_data_ablation}

\begin{figure}[h]
    \centering
    \includegraphics[width=0.9\linewidth]{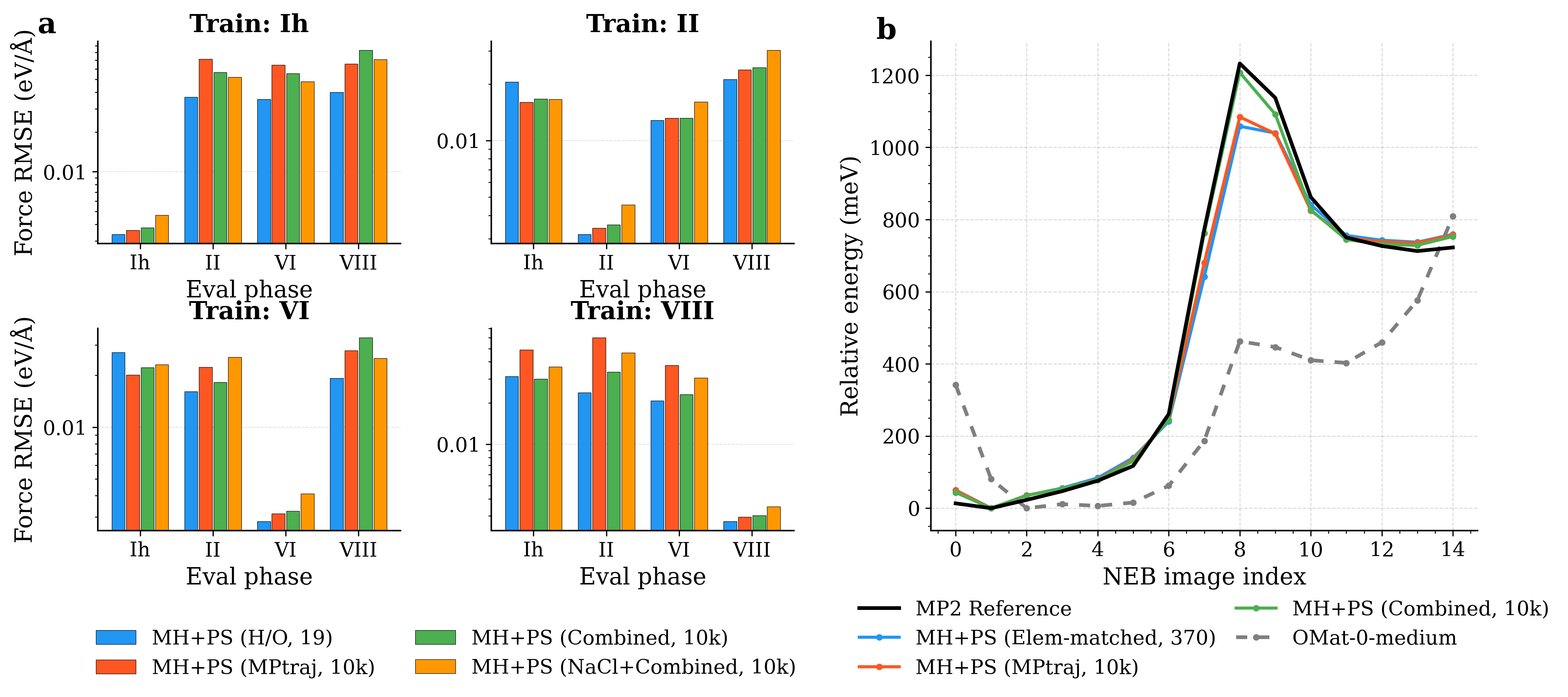}
  \caption{\textbf{Effect of replay-data composition on multihead fine-tuning.} Comparison of three replay-set constructions on the lithium electrolyte system: element-matched subsampling of the OMat24 pretraining corpus, a random 10\,000-structure subsample drawn from MPTraj, and a combination of the two. In all three cases the replay labels are obtained by pseudolabelling the structures with the OMat24 foundation model. Target-task and replay-task force RMSEs are similar across the three choices, indicating that target-task accuracy is largely insensitive to the precise composition of the replay set provided sufficient structural diversity is included.}
      \label{fig:replay_data_ablation}
\end{figure}

We compare three choices for how the replay subsample is constructed for multihead fine-tuning on the lithium electrolyte system: (i) \emph{element-matched subsampling} from the OMat24 pretraining corpus, which retains only structures whose constituent species are a subset of those in the fine-tuning dataset; (ii) a \emph{random 10\,000-structure subsample} drawn uniformly from MPTraj (used because the OMat24 source is not always available, and MPTraj is structurally similar though not identical); and (iii) a \emph{combination}, concatenating the element-matched OMat24 subsample with the random MPTraj subsample. To keep the comparison apples-to-apples across replay sources at different levels of theory, in all three cases the replay labels are obtained by pseudolabelling the chosen structures with the OMat24 foundation model rather than using their original DFT labels. Figure~\ref{fig:replay_data_ablation} shows that target-task accuracy is essentially independent of which of the three choices is used, and that all three preserve foundation-model accuracy on the replay distribution equally well. This insensitivity supports the more general use of pseudolabelled replay: if structural diversity rather than exact provenance of the structures is the relevant property, then the structure source can be decoupled from the label source, and the foundation model itself can supply consistent replay labels for any structurally diverse dataset.

\subsection{NaCl: replay-strategy comparison on PES holes}
\label{app:nacl_replay}

\begin{figure}[h]
    \centering
    \includegraphics[width=0.7\textwidth]{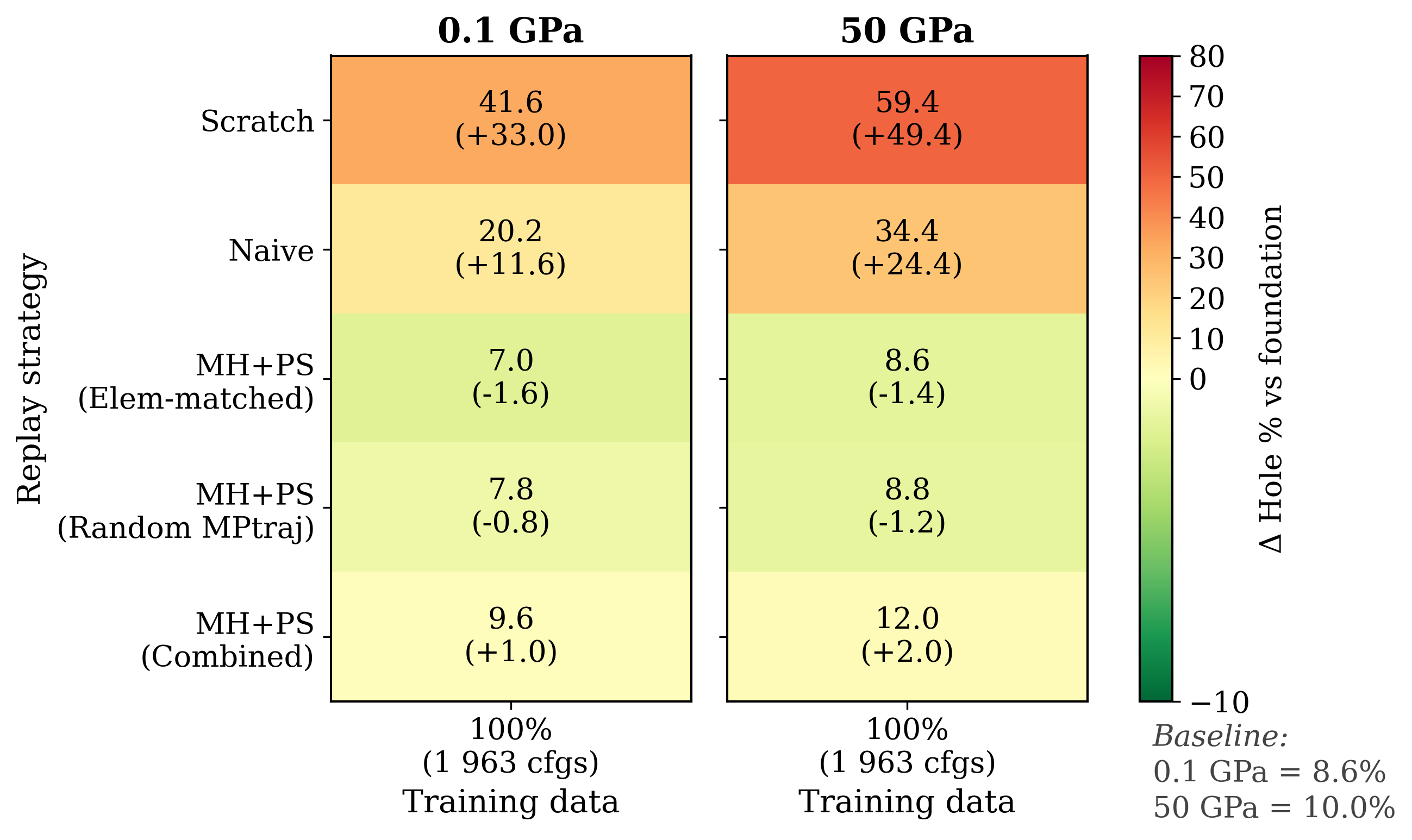}
   \caption{\textbf{Replay-strategy comparison on NaCl PES holes.} Fraction of flagged RSS structures (with absolute change relative to the foundation baseline before fine-tuning shown in parentheses) at 0.1\,GPa and 50\,GPa for Scratch, Naive, and three multihead-pseudolabel replay-set choices (element-matched OMat24 subsample, random MPTraj subsample, combination of the two), all on 100\% of the NaCl training data. All three replay-set choices substantially reduce the hole fraction relative to Scratch and Naive, with only small differences between them.}
     \label{fig:nacl_replay_si}
\end{figure}

To probe how the replay-set choice affects the short-range repulsive wall on the NaCl system, Figure~\ref{fig:nacl_replay_si} compares the fraction of flagged RSS structures across Scratch, Naive, and three multihead-pseudolabel replay-set choices (element-matched OMat24 subsample, random MPTraj subsample, and the combination of the two) on 100\% of the NaCl training data, evaluated at 0.1\,GPa and 50\,GPa. All three replay-set choices reduce the hole fraction substantially relative to Scratch and Naive, and the differences between replay-set choices are small (within $\sim$3\,\% absolute) at both pressures. This is consistent with the replay-data composition comparison in Appendix~\ref{app:replay_data_ablation}: structural diversity in the replay set matters more than the precise provenance of the structures.

\section{Foundation model comparison: foundation baselines and method consistency}
\label{app:foundation_extended}

The main-text Figure~\ref{fig:foundation_model_quality}a shows multihead-pseudolabel fine-tuning across foundation models for compactness. The two figures below extend that comparison in the two directions referenced from the main text: (i) including the foundation model baselines alongside their fine-tuned counterparts, and (ii) comparing multiple fine-tuning methods (foundation baseline, Naive, Multihead-pseudolabel) for each foundation model.

\begin{figure}[h]
    \centering
    \includegraphics[width=0.95\textwidth]{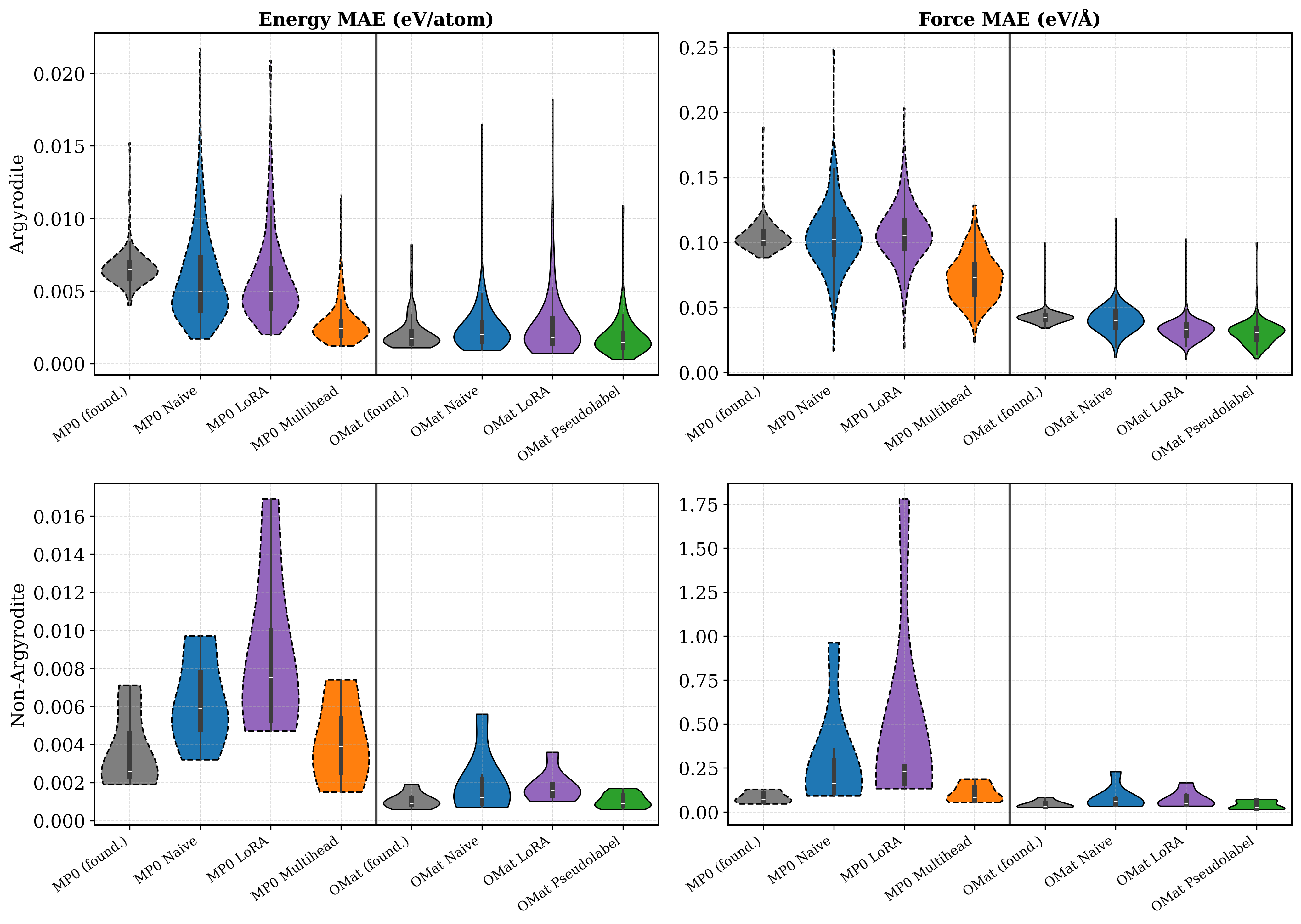}
    \caption{\textbf{Foundation models versus fine-tuning methods on the lithium electrolyte system.} Violin plots of energy MAE (left) and force MAE (right) on the two out-of-distribution evaluation sets used in the main text---other argyrodite compositions (top row) and non-argyrodite structures (bottom row)---for the MP0 and OMat foundation families. Within each foundation, the panels show the foundation baseline before fine-tuning (``found.''), Naive, LoRA, and Multihead/Pseudolabel fine-tuning. The OMat foundation baseline achieves errors comparable to the best fine-tuned MP0 models on both evaluation sets.}
    \label{fig:li_foundation_mh_violin}
\end{figure}

Figure~\ref{fig:li_foundation_mh_violin} reports force and energy MAE on the out-of-distribution argyrodite and non-argyrodite evaluation sets for the foundation baselines before fine-tuning and for Naive, LoRA, and Multihead fine-tuning of each foundation. The MACE-OMat foundation baseline (leftmost grey violin in each OMat panel) already achieves force and energy errors at or below those of the best fine-tuned MPTraj-based models, supporting the statement in the main text that, on this system, foundation quality can outweigh whether fine-tuning is applied.

\clearpage

\section{Foundation model capacity in ice cross-phase learning}
\label{app:ice_capacity}

Foundation model capacity affects in-phase accuracy: MACE-MH1 achieves the lowest force RMSE on the training phase, followed by MACE-OMat-0-medium and MACE-OMat-small. The advantage narrows for cross-phase evaluation, where all models face the same distribution shift. The MACE-MH1 advantage likely reflects both its improved architecture and its more diverse pretraining data (6 datasets vs OMat24 alone); disentangling model capacity from pretraining diversity would require a controlled ablation.

\begin{figure}[h]
    \centering
    \includegraphics[width=0.8\textwidth]{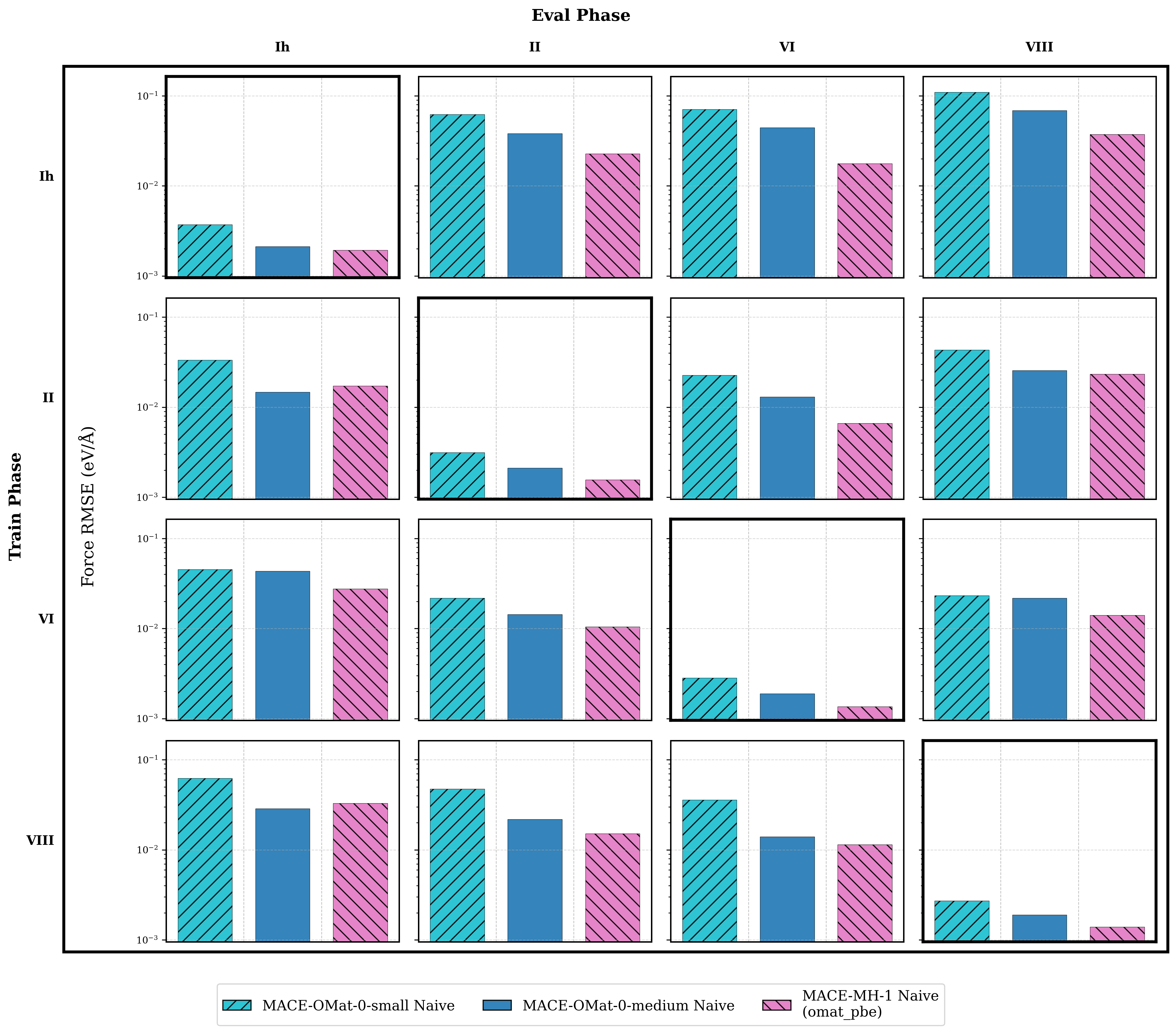}
   \caption{Effect of foundation-model capacity on ice cross-phase learning with naive fine-tuning on 100 configurations. Force RMSE is summarised as in-phase and out-of-phase averages for MACE-OMat-0-medium, MACE-OMat-small, and MACE-MH1. Higher-capacity and more broadly pretrained models achieve lower in-phase errors, though the advantage narrows for cross-phase evaluation.}
     \label{fig:water_cross_foundations_app}
\end{figure}

\subsection{Single-phase versus combined ice fine-tuning}
\label{app:combined_vs_single}

\begin{figure}[h]
    \centering
    \includegraphics[width=0.95\textwidth]{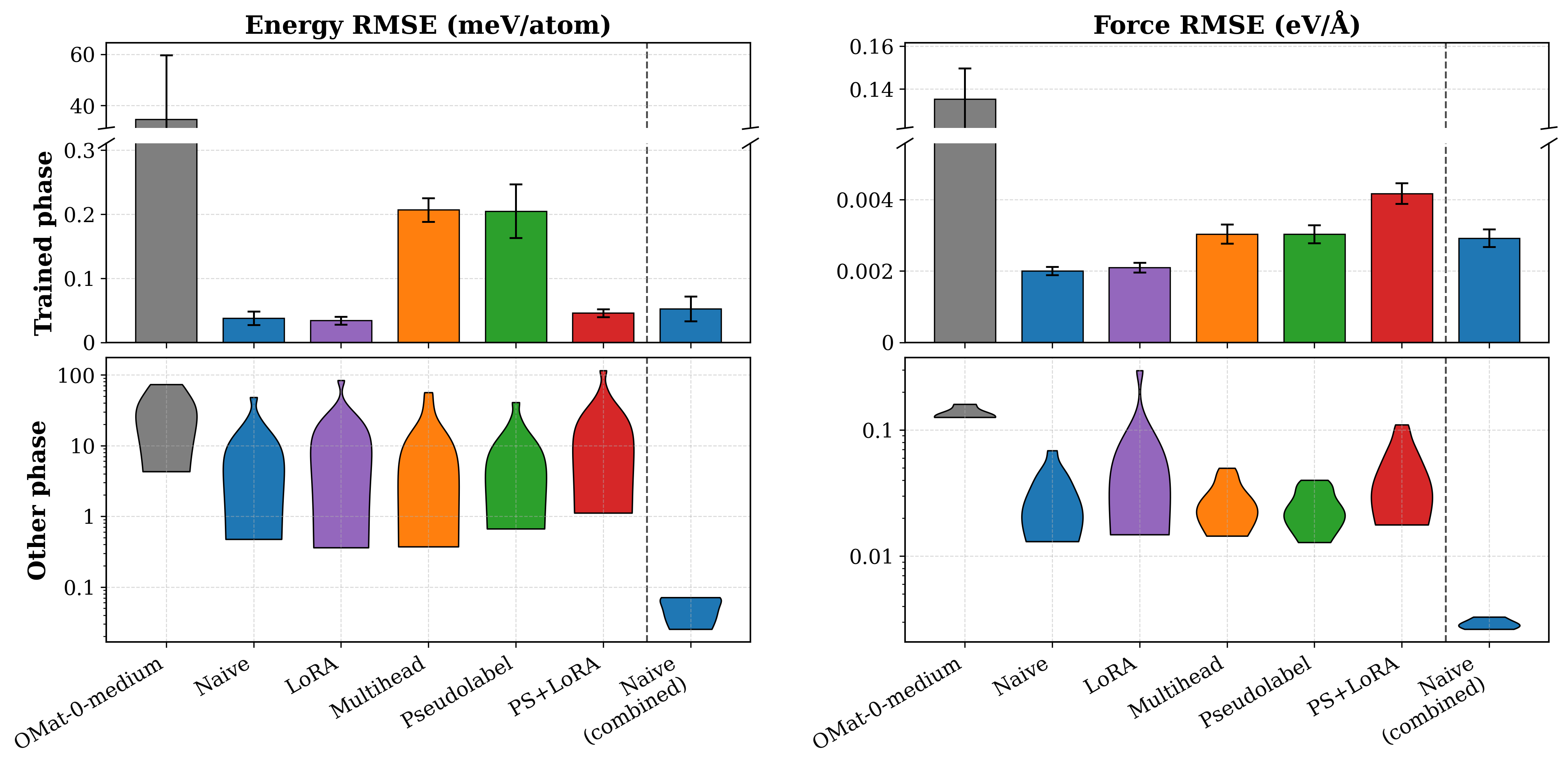}
   \caption{\textbf{Single-phase versus combined ice fine-tuning across methods.} Energy RMSE (left, eV/atom) and force RMSE (right, eV/\AA) for ice Ih, II, VI, and VIII, comparing single-phase fine-tuning with a reference model trained on all four phases jointly. Results are shown for Naive, LoRA, and Pseudolabel fine-tuning of MACE-OMat-0-medium. Dashed vertical lines separate the single-phase evaluation blocks from the combined-phase reference, which is included in each evaluation panel to show how jointly training on all phases changes the cross-phase error scale. All axes use logarithmic scaling.}
     \label{fig:combined_vs_single_si}
\end{figure}

A separate question from foundation-model scale is whether training on a single ice phase or on all four ice phases combined gives better cross-phase performance per method. Figure~\ref{fig:combined_vs_single_si} compares the two protocols across all four ice phases (Ih, II, VI, VIII) for Naive, LoRA, and Pseudolabel fine-tuning of MACE-OMat-0-medium. Combined-phase training does not uniformly outperform the best single-phase model on each evaluation phase: for Naive and LoRA, single-phase training tends to give slightly lower in-phase energy and force RMSEs on phases with sufficient internal diversity (notably Ih), while combined training provides a more uniform error across the four phases. Pseudolabel replay is the most affected by this protocol choice, with combined-phase training yielding noticeably higher errors than single-phase training across all evaluation phases. These results indicate that mixing chemically related but structurally distinct phases into a single fine-tuning set is not automatically beneficial, and that single-phase fine-tuning remains a competitive baseline when the deployment task is restricted to one phase.

\clearpage
\section{NaCl: full RDF and EMD panel}
\label{app:nacl_full}

\begin{figure}[h]
    \centering
    \includegraphics[width=0.95\textwidth]{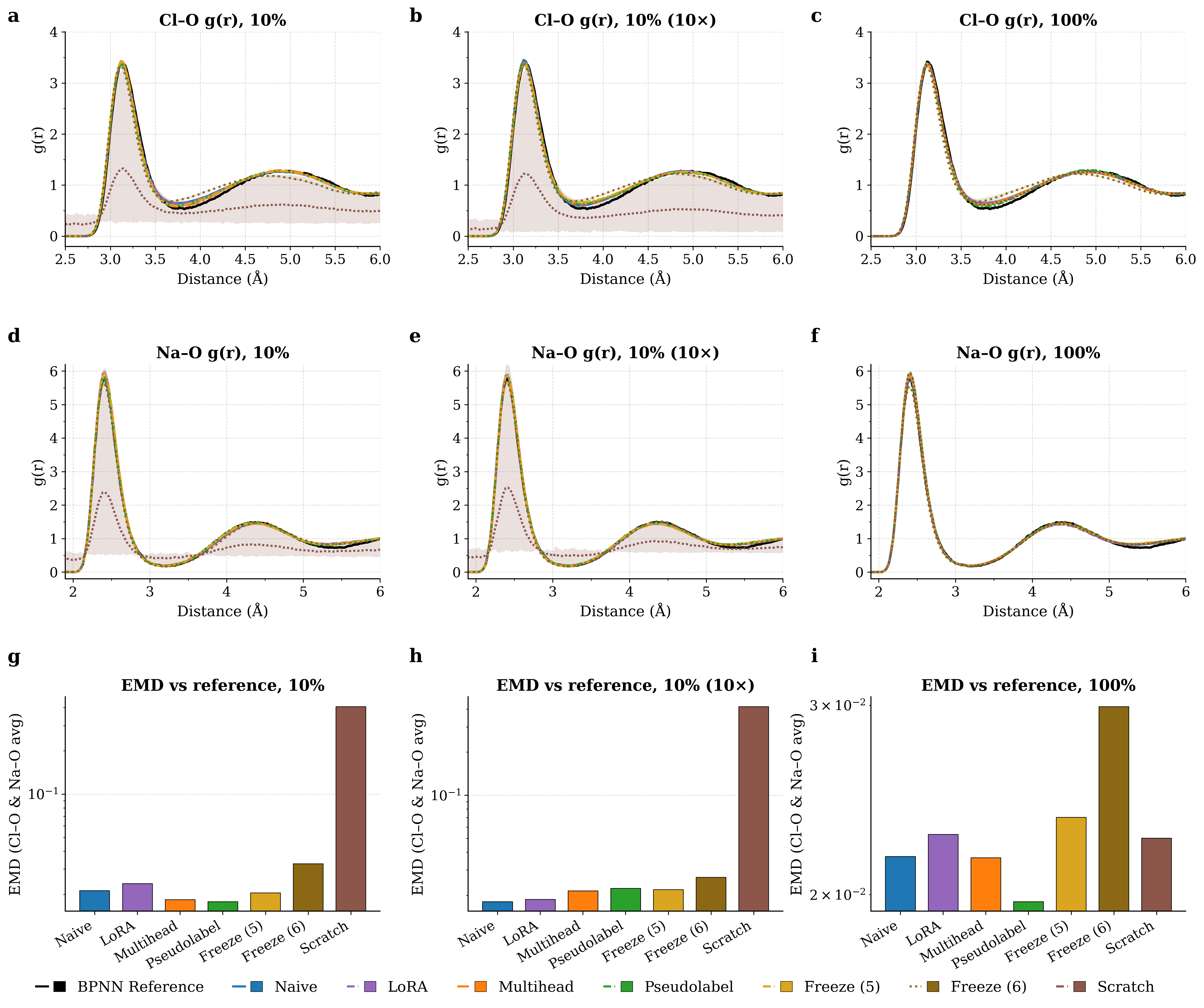}
    \caption{\textbf{Full aqueous-NaCl RDF and EMD panel.} \textbf{a--c}, Cl--O RDFs at 10\%, 10\%-10$\times$ epochs, and 100\% training data. \textbf{d--f}, corresponding Na--O RDFs. \textbf{g--i}, EMD to the BPNN reference for each method at each training-data condition. Fine-tuning methods saturate near the reference already at 10\% data; from-scratch training remains substantially worse even at 100\%.}
    \label{fig:nacl_full_si}
\end{figure}

To complement the 10\%-data RDFs shown in the main text (Fig.~\ref{fig:nacl_rdf_10pct}), Figure~\ref{fig:nacl_full_si} reports the full NaCl fine-tuning panel including the 10\%, 10\%-10$\times$ (extended training), and 100\% training-data conditions, together with the corresponding EMD bar plots for each fraction. The trends discussed in the main text---near-saturation of fine-tuning accuracy already at 10\% data, marginal gains from longer training, and the persistent failure of from-scratch training---are visible across all three data fractions and across both the Cl--O and Na--O RDFs.

\clearpage

\end{document}